\documentclass[]{aastex631}

\usepackage{url}
\usepackage{tabularx}
\usepackage{amsmath}
\usepackage{rotating}
\usepackage{multirow}
\usepackage{xcolor}
\newcommand{\ionwl}[3]{\ion{#1}{#2}~{#3}~{\AA}}

\shorttitle{Can Proton Beam Heating Flare Models Explain Sunquakes?}
\shortauthors{Sadykov et al.}

\begin{document}

\title{Can Proton Beam Heating Flare Models Explain Sunquakes?}

\author[0000-0002-4001-1295]{Viacheslav M. Sadykov}
\affiliation{Physics \& Astronomy Department, Georgia State University, Atlanta, GA 30303, USA}

\author[0000-0002-5519-8291]{John T. Stefan}
\affiliation{Center for Computational Heliophysics, Physics Department, New Jersey Institute of Technology, Newark, NJ 07102, USA}

\author{Alexander G. Kosovichev}
\affiliation{Center for Computational Heliophysics, Physics Department, New Jersey Institute of Technology, Newark, NJ 07102, USA}
\affiliation{NASA Ames Research Center, Moffett Field, CA 94035, USA}

\author{Andrey M. Stejko}
\affiliation{Center for Computational Heliophysics, Physics Department, New Jersey Institute of Technology, Newark, NJ 07102, USA}

\author[0000-0001-7458-1176]{Adam F. Kowalski}
\affiliation{National Solar Observatory, University of Colorado Boulder, 3665 Discovery Drive, Boulder, CO 80303, USA}
\affiliation{Department of Astrophysical and Planetary Sciences, University of Colorado, Boulder, 2000 Colorado Ave, CO 80305, USA}
\affiliation{Laboratory for Atmospheric and Space Physics, University of Colorado Boulder, 3665 Discovery Drive, Boulder, CO 80303, USA.}

\author[0000-0003-4227-6809]{Joel C. Allred}
\affiliation{NASA Goddard Space Flight Center, Solar Physics Laboratory, Code 671, Greenbelt, MD 20771, USA}

\author[0000-0001-5316-914X]{Graham~S. Kerr}
\affiliation{NASA Goddard Space Flight Center, Solar Physics Laboratory, Code 671, Greenbelt, MD 20771, USA}
\affiliation{Department of Physics, Catholic University of America, 620 Michigan Avenue, Northeast, Washington, DC 20064, USA}

\begin{abstract}
    SDO/HMI observations reveal a class of solar flares with substantial energy and momentum impacts in the photosphere, concurrent with white-light emission and helioseismic responses, known as sunquakes. Previous radiative hydrodynamic modeling has demonstrated the challenges of explaining sunquakes in the framework of the standard flare model of `electron beam' heating. One of the possibilities to explain the sunquakes and other signatures of the photospheric impact is to consider additional heating mechanisms involved in solar flares, for example, via flare-accelerated protons. In this work, we analyze a set of single-loop Fokker-Planck and radiative hydrodynamics RADYN+FP simulations where the atmosphere is heated by non-thermal power-law-distributed proton beams which can penetrate deeper than the electron beams into the low atmospheric layers. Using the output of the RADYN models, we calculate synthetic Fe~I~6173\AA~line Stokes profiles and from those the line-of-sight (LOS) observables of the SDO/HMI instrument, as well as the 3D helioseismic response and compare them with the corresponding observational characteristics. These initial results show that the models with proton beam heating can produce the enhancement of the HMI continuum observable and explain qualitatively generation of sunquakes. The continuum observable enhancement is evident in all models but is more prominent in ones with $E_{c}\geq$500~keV. In contrast, the models with $E_{c}\leq$100~keV provide a stronger sunquake-like helioseismic impact according to the 3D acoustic modeling, suggesting that low-energy (deka- and hecto-keV) protons have an important role in the generation of sunquakes.
\end{abstract}


\section{Introduction} \label{sec:introduction}

    The energy release process in solar flares affects all layers of the solar atmosphere, from the photosphere to the corona. The standard `thick-target' flare model assumes that a substantial part of the flare energy is released in the solar corona in the form of a high-energy (deka-keV) electron distribution traveling downward along magnetic field lines and heating the upper chromosphere \citep{Hudson1972}. Radiative hydrodynamic modeling of the atmospheric response to the electron beam heating, initiated by \citet{Kostiuk1975}, revealed upflows of chromospheric plasma into the flare loops (commonly referred to as `chromospheric evaporation'), accompanied by a dense downward propagating shock \citep[chromospheric `condensation',][]{Livshits1981,Fisher1985,Kosovichev1986}. These hydrodynamic models found that the downward-moving shock quickly decays, within 60~s or so, which is borne out by observations \citep[see also ][]{2021ApJ...912...25A,2022ApJ...926..164A,2022FrASS...960856K,2023FrASS...960862K}. This has the implication that the downward-moving shock is insufficient for explaining deep perturbations of the solar photosphere. \citet{Livshits1981} suggested that the white-light emission can be produced by chromospheric condensations \citep[see also][who studied both solar and stellar models of continuum emission from chromospheric condensations]{Kowalski2015b,2017ApJ...836...12K}. The radiative back-warming process can also play an important role in white-light emission generation \citep{Machado1989}.
    
    Using observations from the Michelson Doppler Imager (MDI) on Solar and Heliospheric Observatory \citep{Scherrer1995}, \citet{Kosovichev1998} showed that flares can produce a significant impact in the photosphere, which is sufficient for the generation of helioseismic waves (`sunquakes'). The localized impulsive impacts and sunquakes are also observed in the photospheric observations of the Helioseismic and Magnetic Imager (HMI) onboard the Solar Dynamic Observatory \citep{Scherrer2012}. Such impacts are typically accompanied by enhancements of the continuum emission \citep[for example, all 18 sunquake events studies by][were accompanied by such enhancements]{2015SoPh..290.3151B} as well as strong variations in Doppler shift and magnetic field and are identified as sunquake sources \citep{Sharykin2020}. In the recent work, \citet{Wu2023} suggested that the high-energy tail ($E_{e}>300$\,keV) of a non-thermal electron distribution is a preferred driver of the sunquakes. This result was obtained based on the analysis of 20 strong flares of the Solar Cycle 24 and fitting the hard X-ray emission spectra observed by The Reuven Ramaty High-Energy Solar Spectroscopic Imager \citep[RHESSI,][]{Lin2002}.

     Recently, the flare community has made extensive use of radiative hydrodynamics simulations performed using the RADYN code, developed by \citet{Carlsson1992} and adopted for flare modeling by \citet{Abbett1999} and \citet{Allred2005,Allred2015}. In agreement with the earlier models, these RADYN simulations demonstrated that electron beams can only weakly affect the photospheric layers through direct heating. \citet{sadykov2020FCHROMAruns} performed modeling of the \ionwl{Fe}{1}{6173} Stokes profiles and corresponding SDO/HMI LOS observables for single-loop RADYN electron beam-driven simulations. Those simulations were available as a part of the F-CHROMA\footnote{https://star.pst.qub.ac.uk/wiki/public/solarmodels/start.html} project \citep{Carlsson2023}. The highest HMI continuum intensity observable\footnote{Hereafter when we refer to HMI's continuum intensity observable we drop the `HMI'} enhancement of about 3\%, accompanied by HMI observable Doppler velocities of $\sim$0.4~km~s$^{-1}$, were found for the model with the total energy of $E_{total}$=10$^{12}$~erg cm$^{-2}$, low cut-off energy of $E_{c}=25$~keV, and a power-law spectral index $\delta=3$. The electrons were injected for 20~s into the solar atmosphere in a triangular profile that peaked at t = 10~s for F-CHROMA models, so that correspondingly, the average injected flux in that model was equal to $F_{d} = 5\times{}10^{10}$ erg~cm$^{-2}$~s$^{-1}$. While the perturbations of the SDO/HMI observables in that grid of simulations could not explain the SDO/HMI derived continuum intensity enhancements observed during white light flares \citep[10-100\% depending on the flare's soft X-ray class,][]{song2018WL}, models with higher electron beam fluxes \citep{Kowalski2022} are expected to show a significant white light emission. The F-CHROMA models also clearly do not result in the velocity signals of several km~s$^{-1}$, which are needed for the initiation of sunquakes \citep{Stefan_SQ}.
    
    One of the possibilities to explain the deep perturbations of the photosphere, including sunquakes, is to consider additional heating mechanisms involved in solar flares, for example, Alfv\'en wave heating \citep{2016ApJ...818L..20R,kerr2016alfven} and heating by non-thermal proton beams \citep{prochazka2018protons}. It is likely that non-thermal protons are present in flares, and may even carry energy equivalent to that of the non-thermal electron distribution \citep[][]{Emslie2012}. However, largely owing to poor constraints on the properties of the distribution they are often ignored in flare modelings, and numerical studies of their role in the Sun's atmospheric response to flares are rare. For further discussion see the introduction to \cite{Kerr2023}.

    In this work, we analyze single-loop RADYN proton beam simulations in a wide set of beam parameters and impose the pressure perturbations from these models into the 3D helioseismic model of sunquakes \citep{Stefan_SQ}. Our goal is to answer the question: \textit{can single-loop RADYN proton heating simulations, coupled with the 3D acoustic model of the Sun, cause a photospheric impact and explain the initiation and propagation of helioseismic signals detected in sunquakes?} The paper is structured as follows. Section~\ref{sec:radyngen} describes RADYN proton beam heating simulations, synthesis of the \ionwl{Fe}{1}{6173} line profiles and SDO/HMI line-of-sight (LOS) observables for these simulations, and the related results. Section~\ref{sec:acousticgen} describes the development of the acoustic model simulations driven by the perturbations imposed from the RADYN models and the related results. The summary of the paper results followed by discussion is presented in Section~\ref{sec:discussion}.

\section{Modeling \ionwl{Fe}{1}{6173} Stokes Profiles and SDO/HMI Observables from RADYN Proton Beam Simulations} \label{sec:radyngen}

    \subsection{Description of RADYN Proton Beam Heating Runs} \label{subsec:radyndescription}

        We employ the unified computational model for solar and stellar flares, RADYN \citep{Allred2015}, which self-consistently combines the equations of radiation transport, non-equilibrium atomic level populations, charge conservation, and hydrodynamics following the injection of flare energy into one leg of a semi-circular loop, assumed to be symmetrical. RADYN was recently coupled with the FP code \citep{Allred2020} which models the transport of high-energy particles from their injection in the solar corona to thermalization in the low atmosphere by solving the Fokker-Planck equation. The FP code models the evolution of the distribution function taking into account Coulomb collisions, nonuniform ionization, magnetic mirroring, the return electric currents, and synchrotron emission reactions, and offers an improvement upon prior implementations of non-thermal particles in RADYN. FP implements Coulomb collisions of non-thermal particles with an ambient plasma of arbitrary temperature and is applicable throughout the cold-target to warm-target regimes. Importantly for this study, for non-thermal protons with energies in the deka- to hecto-keV range, the ambient plasma is typically a warm-target, which is much less effective at slowing the protons than cold-target collisions \citep{1986ApJ...309..409T}.

\begin{figure}[h]
\centering
    \includegraphics[width=1.0\linewidth]{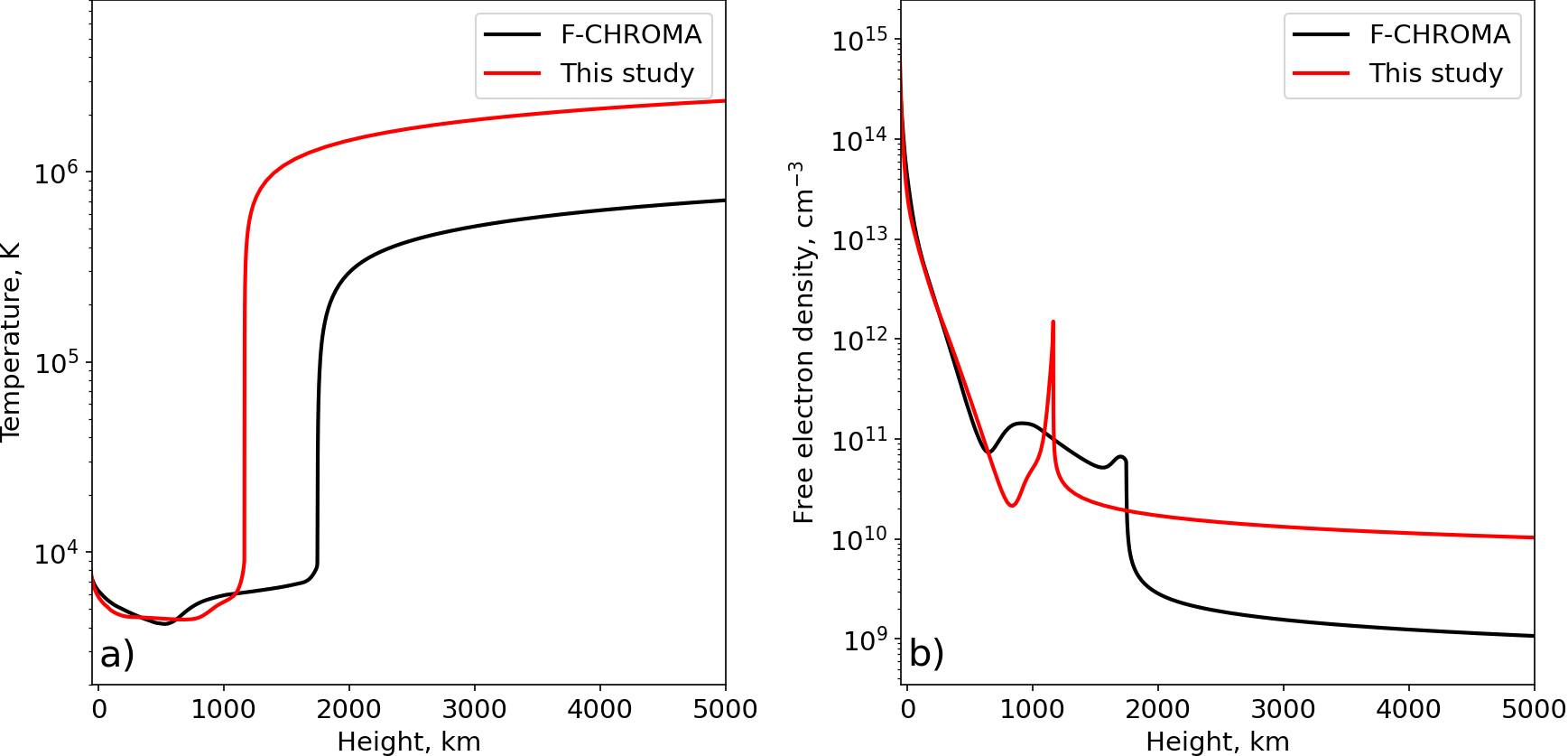}
    \caption{Comparison of the (a) temperature profiles and (b) free electron density profiles for the initial atmospheres used for RADYN simulations in F-CHROMA grid \citep{Carlsson2023} and in this study.}
\label{fig:initatmo}
\end{figure}

        In this work, we take advantage of the proton beam heating simulation grid computed for the series of studies recently started by \citet{Kerr2023}. All RADYN proton beam runs analyzed in this work have approximately the same energy flux rates, $F_{d} = 10^{11}$ erg~cm$^{-2}$~s$^{-1}$, with two values for the energy power-law spectral indexes: $\delta=3$ and 5, and eight values for the low-energy cutoff, $E_{c} =$ 25~keV, 50~keV, 100~keV, 150~keV, 250~keV, 500~keV, 1000~keV, and 3000~keV. The proton beam was injected into the atmosphere during the first 20~s of each simulation, after which the atmosphere evolved for an additional 80~s. Of these 16 total runs 15 are considered here. The run, parameterized by $F_{d} = 10^{11}$ erg~cm$^{-2}$~s$^{-1}$, $\delta = 5$, and $E_{c} = 25$~keV, required a very small time step due to large perturbations of the atmosphere driven by beam heating, and was performed for a shorter duration. The overall properties of the completed runs are summarized in Tables~\ref{tab:stats_f11_d3}~and~\ref{tab:stats_f11_d5}. Note that the pre-flare atmosphere used in the F-CHROMA grid of simulations is rather different than that used here. The pre-flare atmosphere used in the F-CHROMA database is a VALC-like atmosphere, with a 1\,MK corona and chromospheric temperature plateau. The chromospheric structure is maintained by non-radiative heating added to the simulation when beam heating is not present, and its construction is described in \citet{Carlsson2023}. The pre-flare atmosphere used in \citet{Kerr2023} and our study of proton beam heating is an atmosphere in radiative equilibrium such that additional heating is only added to maintain the corona and photosphere. It has an apex temperature of 3\,MK and is denser than the VALC-like atmosphere, with a transition region located at a deeper geometric altitude. \citet{Allred2015} describes the construction of this atmosphere. The difference between the vertical profiles of the initial atmospheres is illustrated in Figure~\ref{fig:initatmo}. In addition, the energy flux is injected at a faster rate than for the F-CHROMA grid, which is injected more gradually over a 20~s triangular pulse, and the energy flux is twice larger on average.

    \subsection{Modeling of \ionwl{Fe}{1}{6173} Stokes Profiles and SDO/HMI Observables} \label{subsec:linemodeling}

        The SDO/HMI obtains the line-of-sight (LOS) observables by measuring the \ionwl{Fe}{1}{6173} line in two polarizations (Right-Circular Polarization, RCP, and Left-Circular Polarization, LCP) over six wavelength points. To model how SDO/HMI would see the flare atmospheres heated by the proton beams, we synthesize the \ionwl{Fe}{1}{6173} Stokes profiles for the RADYN simulations using a similar approach as described in \citet{sadykov2020FCHROMAruns}. In brief, the NLTE Fe atomic level populations and resulting emission from bound-bound transitions are calculated under the assumption of statistical equilibrium using the RH1.5D radiative transfer code \citep{pereira2015RH15D,rybicki1991RHI,rybicki1992RHII,uitenbroek2001RHPRD}. Since RH15D solves the statistical equilibrium equations, non-equilibrium effects are not included. The dynamic hydrogen populations and the free electron densities are imported from the RADYN model and not recalculated, somewhat mitigating the requirement to assume statistical equilibrium in the solution of \ion{Fe}{1}. The full Stokes profiles are solved for the \ionwl{Fe}{1}{6173} transition, with no effects of the background polarization taken into account, and the LCP and RCP profiles are computed. The models are augmented with various settings of the imposed vertical magnetic field unchanged during the simulations. Although we consider a 500~G magnetic field setting (selected for illustration purposes only), the choice of the magnetic field does not impact the hydrodynamic simulation results, and only affects the calculations of the Stokes profiles and HMI observables. The Stokes profiles are computed from RADYN run output that is typically at a cadence of 0.1~s.
        
        The properties of the \ionwl{Fe}{1}{6173} line are computed (1) directly from the fully-resolved Stokes profiles, (2) by applying the SDO/HMI Line-of-Sight (LOS) pipeline to the Stokes profiles and assuming that the polarization signals are obtained instantly for every wavelength/filter, and (3) by applying the SDO/HMI Line-of-Sight (LOS) pipeline to the Stokes profiles following the proper temporal sequence of polarization measurements \citep{schou2012HMImeasurements}. For the properties directly estimated from the fully-resolved Stokes profiles, we compute the continuum intensities (calculated as the average intensity at $\pm$1\,\AA\ from the reference wavelength of the line, $\lambda{}_{ref}$=6173.34\,\AA), line depths (defined as the strongest deviations from the continuum intensities), and line Doppler shifts \citep[calculated using the center of gravity approach; e.g.,][]{Sadykov2019}. The same observables are obtained using a simplified version of the HMI data analysis algorithm following the original method described by \citet{couvidat2012HMI,couvidat2016HMI}. The simplified pipeline is described in detail in \citet{sadykov2020FCHROMAruns} and its application to RADYN electron beam heating runs is tested therein. Here, we provide a brief summary of the pipeline.

        The LCP and RCP profiles of the \ionwl{Fe}{1}{6173} are sampled in six wavelength points by assuming a Gaussian-like transmission profile in the wavelength space (with FWHM$\sim$76\,m\AA). The wavelength points are centered at the rest wavelength of \ionwl{Fe}{1}{6173} ($\lambda{}_{ref}$=6173.34\,\AA) and are sampled at $68.8\,m{\AA}$ apart. From these six measurements, the first and second Fourier components are computed and used for the estimation of the HMI observables \citep[continuum intensity, line depth, Doppler velocity, and LOS magnetic field; see][]{couvidat2012HMI} with the corresponding correction factors \citep{couvidat2016HMI}. The \ionwl{Fe}{1}{6173} line profile is assumed to be Gaussian. In contrast with the previous work \citep{sadykov2020FCHROMAruns}, the line width is not assumed to be fixed but recalculated for every application of the SDO/HMI pipeline. Keeping the line width fixed and equal to the unperturbed line profile width does not change the results and conclusions qualitatively. We sample the line profiles at each time snapshot and compute `instantaneous observables' (i.e., observables where the temporal sequence of measurements is not taken into account).
        
        The HMI LOS pipeline measurement sequence takes 45\,s to be completed. Given that the heating phase lasts only 20\,s, the observables will depend on the time of the heating with respect to the measurement sequence timing. In this work, for every time moment of the simulation, we assume that the SDO/HMI pipeline is centered temporally at that time moment, and compute the observables. Therefore the dynamics of the observables presented further need to be interpreted as what `can possibly be' observed by HMI during the proton beam heating event rather than what 'is' observed. That is, the best-case scenario is HMI happens to catch the start of the impulsive heating of a single pixel. The temporal sequence of polarizations and wavelengths are assumed following \citet{schou2012HMImeasurements}. The polarization profiles for $t<0$\,s are assumed to be the same as the profiles of an unperturbed atmosphere at $t=0$\,s, and the same as for $t=100$\,s time moment for any $t>100$\,s. This allows one to compute the observables during the heating phase and for the last 20\,s of the run.

    \subsection{Results from RADYN} \label{sec:resultsRADYN}

        The summary of the strongest perturbations (both the physical properties of the atmosphere and the properties of the \ionwl{Fe}{1}{6173} line formation and appearance) for all considered RADYN runs is presented in Tables~\ref{tab:stats_f11_d3}~and~\ref{tab:stats_f11_d5}. The perturbations of the atmospheric parameters are computed for the heights of 100~km and 300~km which correspond to the heights where the \ionwl{Fe}{1}{6173} line is typically formed for the quiet Sun conditions \citep{norton2006lines,kitiashvili2015feIsim}. We note here that during the flare process, the \ionwl{Fe}{1}{6173} may experience a significant chromospheric contribution~\citep{Monson2021} which we do not investigate in this work. The height of 100~km is chosen for analysis instead of the 0~km because the latter is close to the bottom boundary of the modeling domain (which is just $\sim$60--65~km below the 0~km height) and affected by the boundary conditions. The selection of the upper height of $h = 300$~km is consistent with maximum values of $\tau_{\lambda} =1$ modeled for the continuum near the \ionwl{Fe}{1}{6173} line presented in the tables (although, as noted earlier, this does not preclude the possibility of the contribution of higher levels of the atmosphere to the \ionwl{Fe}{1}{6173} formation). The atmospheric perturbations at $h = 300$~km for the presented runs range from several percent for the considered parameters (for moderate $E_{c}\sim$ 250~keV values) to several tens of percent for the extremely high or extremely low values of the $E_{c}$. The perturbations are mostly several percent at the height of $h = 100$~km for these runs and rarely reach tens of percent. Vertical velocities rarely reach as much as several hundreds of m s$^{-1}$ at $h = 100$~km.  At that height, the largest downward velocity\footnote{Here we define negative velocities are downflows, and positive velocities are upflows} is in Model 9 with $v_{z}\sim{}-0.794$~km s$^{-1}$. The notation $v_{z}$ hereafter corresponds to the hydrodynamic velocities in RADYN simulations. However, vertical velocities are significantly faster at a height of $h = 300$~km, where, in Model 9, the peak downward velocity is $v_{z}\sim{}-2.78$ km s$^{-1}$. The electron beam-induced velocities for the F-CHROMA grid were significantly lower \citep{Monson2021,sadykov2020FCHROMAruns}. For the atmospheric parameters (temperatures, pressures, and vertical velocities at the heights of $h=100$\,km and $h=300$\,km), the strongest perturbations occur for the lowest and highest values of the low-energy cutoff parameter, $E_{c}$, corresponding to the cases of the highest number of protons or the highest energy per proton, respectively.

        Tables~\ref{tab:stats_f11_d3}~and~\ref{tab:stats_f11_d5} also present the spectroscopic line parameters as computed from the fully-resolved Stokes profiles (i.e. with no application of the SDO/HMI LOS algorithm at this point). The strongest perturbations of the \ion{Fe}{1} line profiles (in terms of their continuum intensities, line depths, and Doppler shifts) and changes to the $\tau{}=1$ heights occur in the cases of very high or very low $E_{c}$. In particular, the continuum intensity enhancement reaches more than 40\% and the strongest blueshift (representing an upflow) reaches almost $v_{D}\sim{}0.97~$km~s$^{-1}$ for Model 8 ($E_{c}=3000$~keV, $\delta =3$), which has the highest average energy per proton considered. As mentioned previously, Model 9 ($E_{c} = 50$~keV, $\delta = 5$) has the strongest redshift of $v_{D}\sim{}-2.78~$km~s$^{-1}$ at $h = 300$~km.

        Figure~\ref{fig:stats} illustrates the maximum values of the relative continuum intensity enhancements, $I_{c}/I_{c}^{0}$, and the strongest redshift, min\,$v_{D}$ (normally interpreted as a downflow speed, though opacity effects make a one-to-one relation difficult), as functions of the $E_{c}$ and $\delta$ parameters of the proton beam energy spectrum. Hereafter, $v_{D}$ corresponds to the velocities inferred from Doppler shifts of fully-resolved Stokes profiles applying a center of gravity approach. While both the continuum intensity enhancement and strongest redshifts tend to increase with the change of the $E_{c}$ for both $\delta$ setups, the trends for $I_{c}/I_{c}^{0}$ and $v_{D}$ are slightly different. The enhancement of the continuum near the \ionwl{Fe}{1}{6173} spectral line is in the range of 5-10\% for most of the runs and increases consistently with the increase of $E_{c}$. The situation is the opposite for Doppler shifts, where the strongest redshifts tend to increase with the decrease of $E_{c}$.

        Among the considered models, two are of a special interest. Model 8 has the strongest enhancement of the continuum intensity near the line, and Model 9 has the strongest perturbation of the Doppler shift $v_{D}$ of the line profile. Therefore, both models are chosen for more detailed analysis. Figure~\ref{fig:atmosdynamics} displays the evolution of the temperature and pressure relative to their unperturbed values (at $t=0$~s) and the vertical velocities for these two models. One can see that both models resulted in significant perturbations of the lower atmosphere. Figure~\ref{fig:atmosmodels} visualizes the behavior of temperature, vertical velocity, gas pressure, and the $\tau{}=1$ height of the \ionwl{Fe}{1}{6173} line core and nearby continuum as a function of time. Model 8 has a stronger temperature response and a larger change in the $\tau{}=1$ heights than does Model 9. In Model 9, during the time interval of $\approx$30--90~s, there is extensive enhanced pressure in the region of line formation that is not evident in Model 8. In contrast to Figure~\ref{fig:atmosdynamics}e, Figure~\ref{fig:atmosdynamics}b shows the downward propagation of the pressure enhancement peak in the lower atmosphere, below 750\,km. In contrast, for Model 8 the increase of the gas pressure happens during the first 20~s of the run during the heating phase.
        
        \subsection{Analysis of Stokes profiles}

        Figures~\ref{fig:profiles1}~and~\ref{fig:profiles2} illustrate the evolution of the \ionwl{Fe}{1}{6173} left- and right-circular polarization profiles (LCP and RCP) throughout the simulation. Overall, the line profiles presented in Figure~\ref{fig:profiles1} (corresponding to Model 9) do not experience notable changes until after the initial heating phase, when they become asymmetric: the blue wing of the profiles deepens between $t\approx$20--30~s and then enhances, with the strongest enhancement happening at around t$\sim$60~s. Figure~\ref{fig:atmosmodels}b demonstrates that the strongest enhancement of the blue wing follows the strongest downward motions of the atmosphere at $h = 300$~km by just several seconds. In contrast to that behavior, the dynamics in Model 8 result in more enhanced, complex shapes of the line profiles during the heating phase and right after it, as seen in Figure~\ref{fig:profiles2}a-c. In particular, the \ion{Fe}{1} line LCP and RCP profiles demonstrate an emission feature superimposed with the absorption and have an enhanced blue wing after 10~s of the start of the run. The enhancement of the \ion{Fe}{1} polarization profiles is also reflected in $\tau{}=1$ heights: the heights lie in a more shallow region, within $h\sim$50--200~km (see Figure~\ref{fig:atmosmodels}h). 
        
        In addition to properties derived directly from the LCP and RCP profiles, we have applied the simplified SDO/HMI pipeline \citep[][also described in Section~\ref{subsec:linemodeling}]{sadykov2020FCHROMAruns} and computed the instantaneous and time-dependent line-of-sight (LOS) observables. The results are presented in Figure~\ref{fig:timedep} for the case of a 500~G imposed vertical magnetic field, and do not change qualitatively for other values of the field. The instantaneous observables demonstrate the same patterns as the spectral line properties (continuum intensities near the line, line depths, and Doppler shifts) derived from the full-resolution line profiles for Models 8 and 9. The systematic offset of the synthetic observables with respect to the quantities derived from their full-resolution counterparts is typically a result of the non-Gaussian shape of the \ion{Fe}{1} line (see Figures~\ref{fig:profiles1}~and~\ref{fig:profiles2}) while the SDO/HMI LOS pipeline relies on the Gaussian-shape assumption~\citep{couvidat2016HMI}. The behavior of the time-dependent observables (taking into account the timing of the polarization measurements at different wavelengths) is close to the instantaneous observables for Model 9 (Figure~\ref{fig:timedep}a-d). However, for Model 8 the behavior of the line depth (Figure~\ref{fig:timedep}f) and Doppler velocity (Figure~\ref{fig:timedep}g) observables differ dramatically. In particular, while the redshifts measured from the fully-resolved LCP and RCP profiles reach only $\approx$-0.7~km~s$^{-1}$, the time-dependent SDO/HMI observables demonstrate Doppler shifts greater than -2~km~s$^{-1}$. The vertical magnetic field observable fluctuates for this model approximately in the range of 300-900~G, while its true value was always unchanged and equal to 500~G. Such dynamics are due to the strong impulsive heating whose duration (20~s) is less than the duration of the SDO/HMI LOS pipeline procedure (45~s). Model~9~(Fig.~\ref{fig:timedep}a-d) shows a much better agreement between the properties of the line profile derived directly from the spectrum and the SDO/HMI observables with respect to Model~8~(Fig.~\ref{fig:timedep}e-h), most likely because of the more gradual evolution of the \ion{Fe}{1} polarization profiles evident in Figures~\ref{fig:profiles1}~and~\ref{fig:profiles2}. Overall, it once again confirms that SDO/HMI LOS observables, including a continuum intensity observable, have to be interpreted with caution during solar flares \citep{Svanda2018}.

\section{3D Acoustic Models Driven by RADYN Simulations}
\label{sec:acousticgen}

    In this section, we use the atmospheric response to proton beam heating---as computed by RADYN---as an input to an acoustic model to measure the amplitude of helioseismic waves that may be generated. We provide a brief description of the acoustic model and the coupling procedure for the RADYN output in Section \ref{sec:acoustic model description} followed by the results of this modeling in Section \ref{sec:acoustic model results}.

    \subsection{Model Description}\label{sec:acoustic model description}

        The 3D acoustic model treats acoustic oscillations as linear, adiabatic perturbations to pressure, density, and velocity \citep{Stefan_SQ}. The background stratification for the solar interior is derived from the Standard Solar Model \citep{SSM} which smoothly transitions to the atmosphere used in the RADYN simulations. The transition location in the Standard Solar Model, at $R=695.707$~Mm relative to the solar center, is chosen where the mass density is equal to that at $z=0$~km in the RADYN mesh.

        The computational acoustic model itself is semi-spectral, with the radial derivatives evaluated numerically and polar- and azimuthal-angle derivatives evaluated spectrally using spherical harmonics. While the acoustic model does not include radiative damping, the generated oscillations are damped according to the horizontal wavenumber, with damping parameters derived from quiet-Sun pressure wave (p-mode) data reported by \citet{pmodefreq}. We consider the choice of quiet-Sun damping parameters, as opposed to active region-like, appropriate here as the majority of the sunquake wavefront propagates outside the generating active region, where the magnetic field is moderate or weak.
        
        We take advantage of the advanced treatment of radiation in RADYN by deriving our acoustic model input using the simulated perturbations to gas pressure. These gas pressure perturbations are smaller in RADYN than they would be in the acoustic model as RADYN accounts for the energy lost in optically thin and NLTE optically thick radiation. Here, gradients in the gas pressure determine the acceleration a plasma parcel experiences from the supplied heating, and the input accelerations are computed from the RADYN simulations with
        $\dot{v_r} = (1/\rho_{0})\cdot{\partial P^{\prime}}/{\partial r}$,
        where $\rho_0$ is the initial state's mass density. An example of the input acceleration profiles, as described in the next section, for low-energy cut-offs $E_c = 50$~keV and $E_c = 3000$~keV (Models 9 and 15) interpolated onto the hydrodynamic model's background mesh is shown in Figure \ref{fig:acc}. We highlight these models in particular as they span the two extremes of our low-energy cut-off parameter space.

        We assume a constant cross-sectional area, and the horizontal profile of the input accelerations is considered to be Gaussian with an FWHM of 1500\,km. The FWHM is based on HMI observations of sunquake kernels that range in size from one to several pixels; this corresponds to an impact site between 750\,km and 2000\,km. Where the lower end of the supplied accelerations ends, slightly below $R=695.707$~Mm, the functions are appended by a Gaussian with a drop-off closely matching the unappended input. A similar drop-off is applied to the upper end of the supplied accelerations, which extend to the top of the modeled corona, to avoid boundary effects that may be caused by providing input close to the upper boundary. The Gaussian upper drop-off begins 350 km from the upper boundary with an FWHM of 125~km. An additional description of the acoustic model, including numerics, is provided in \cite{Stefan_SQ}.

    \subsection{Results from the Acoustic Model}\label{sec:acoustic model results}

        We are primarily interested in the behavior of the photospheric radial velocity, as this is generally the largest component of the line-of-sight velocity in SDO HMI Dopplergrams for observations close to the disk center. We then examine the resulting photospheric p-mode wavefront, with absolute maximum radial velocity for each case shown as a function of horizontal distance from the beam target in Figure \ref{fig:amp}. Note that acoustic-gravity waves---not typically observed in actual sunquake events---are generated in addition to the usual p-mode wavefront. In the cases where the cut-off energy is greater than 250~keV, these acoustic-gravity waves have amplitudes that exceed the p-mode wavefront. We, therefore, examine the absolute maximum amplitude at each distance within 5 minutes of the wavefront travel time predicted by ray theory; however, it is not possible to disentangle the two wavefronts for short ($<$5--7~Mm) distances.

        We observe core velocities in the acoustic model which are significantly greater than those in the RADYN simulations, in particular for the simulations with low cut-off energies. For example, the acoustic model predicts a magnitude of the radial velocity for Model 9 ($E_c=50$~keV, $\delta=5$) of 68.7~km~s$^{-1}$; in such cases, we do not expect the predictions that are close to the source to be reliable because of both the linear nature of the model and the lack of radiative damping. Conversely, the magnitude of the radial velocity for Model 15 ($E_c=3000$~keV, $\delta=5$) at the beam core is only 0.751~km~s$^{-1}$ which securely falls in the linear regime. Thus we consider the entire range of the acoustic model to be reliable for the higher cutoff-energy cases.

        The relationship between the absolute maximum radial velocities of each case remains largely the same over distance as compared to the beam core. In general, the greatest absolute velocities are associated with small low-energy cut-offs with the wavefront amplitudes decreasing with increasing low-energy cut-off. There is relatively little deviation in this relationship when increasing the spectral index from $\delta=3$ to $\delta=5$, though we note that the radial velocities for the $E_c=500$~keV cases do change appreciably from the $\delta=3$ case (Model 6) to $\delta=5$ case (Model 13). In the associated acceleration profile from each model, we observe significantly stronger evaporation in the $\delta=5$ case as well as a downward-propagating acceleration front that penetrates slightly more deeply in Model 13 than in Model 6. While the initial evaporation in each case is similarly impulsive---that is, the evaporation fronts propagate upwards with similar speeds---more of the beam energy is released in the initial evaporation of Model 13. For models with lower energy cut-offs, there is a similar increase in acceleration magnitude when moving from $\delta=3$ to $\delta=5$. The downward-propagating acceleration front is weak for low-energy cut-off $E_c=1000$~keV (Models 7 and 14) and non-existent for low-energy-cutoff $E_c=3000$~keV (Models 8 and 15), and the radial velocity in these two cases changes the least with spectral index. The lack of the downward-propagating front in Model 15 is clearly seen in the comparison with Model 9 in Figure~\ref{fig:acc}.

        We now look more closely at the maximum radial velocity at a horizontal distance of $X=18$~Mm indicated by the vertical dashed line in Figure \ref{fig:amp}, where the sunquake wavefront is expected to reach its maximum amplitude (aside from the beam core), based on observations (cf. Figure 3d in \citet{Macrae2018}, Figure 7 in \citet{Zharkov2020}, and Figure 2 in \citet{Sharykin2020}). Explicitly plotting these velocities on a log-log scale, as in Figure \ref{fig:18Mm}, we find that the generated sunquakes fall into two separate regimes. There is a low-energy cut-off regime extending up to $E_c=250$~keV where the sunquake amplitudes are similar to observations (on the order of 100s of m~s$^{-1}$), and a high-energy cut-off regime beginning at $E_c = 1000$~keV where the sunquake amplitude is significantly lower than in observations.

\section{Summary and Discussion} \label{sec:discussion}

\begin{figure}[h]
\centering
    \includegraphics[width=1.0\linewidth]{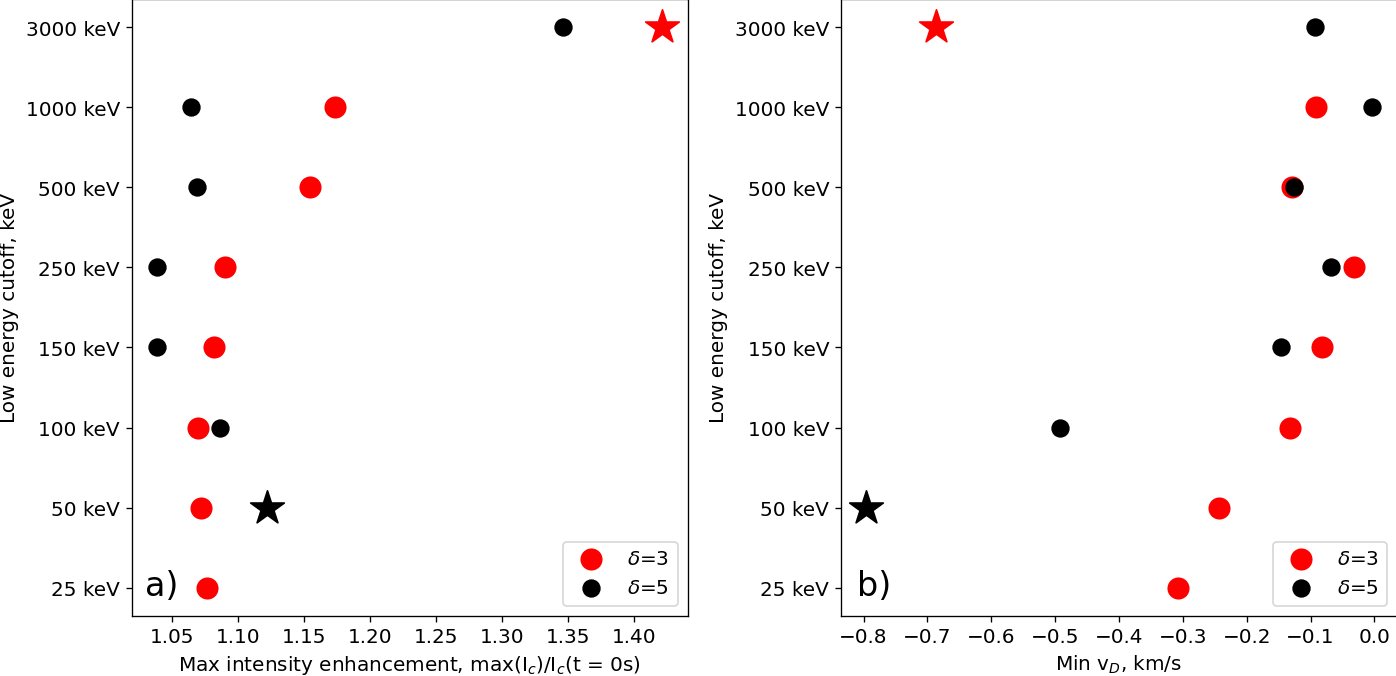}
    \caption{The maximum intensity enhancement (a) and the strongest Doppler redshift (b) as a function of the low-energy cutoff, $E_{c}$, for the considered models. The visualization is based on the RADYN model summaries in Table~\ref{tab:stats_f11_d3}~and~\ref{tab:stats_f11_d5}. The models selected for the detailed analysis (Model 8, $E_{c} = 3000$~keV, $\delta =3$, $F_{d} = 10^{11}$ erg~cm$^{-2}$~s$^{-1}$, and Model 9, E$_{c} =50$~keV, $\delta = 5$, $F_{d} = 10^{11}$ erg~cm$^{-2}$~s$^{-1}$) are marked by asterisks.}
\label{fig:stats}
\end{figure}

\begin{figure}[h]
\centering
    \includegraphics[width=1.00\linewidth]{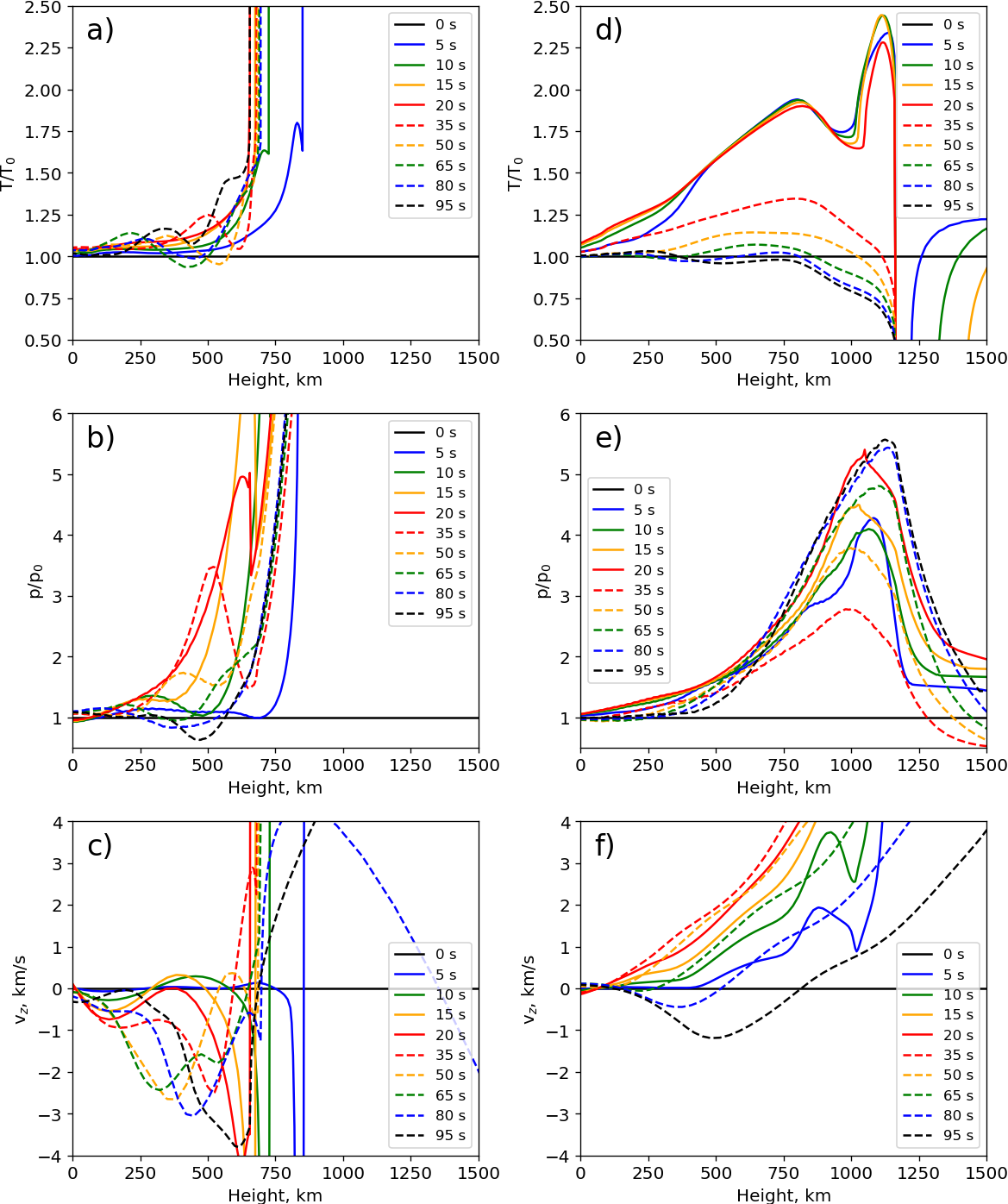}
    \caption{Evolution of the temperature enhancement relative to t=0\,s moment (a), gas pressure enhancement relative to t=0\,s moment (b), and vertical velocity (c) at the different time moments for RADYN Model 9 (E$_{c} =50$~keV, $\delta = 5$, $F_{d} = 10^{11}$ erg~cm$^{-2}$~s$^{-1}$). Panels (d-f) show the same parameters as in panels (a-c) but for the RADYN Model 8 ($E_c = 3000$~keV, $\delta = 3$, $F_{d} = 10^{11}$~erg~cm$^{-2}$~s$^{-1}$.)}
\label{fig:atmosdynamics}
\end{figure}

\begin{figure}[h]
\centering
    \includegraphics[width=0.49\linewidth]{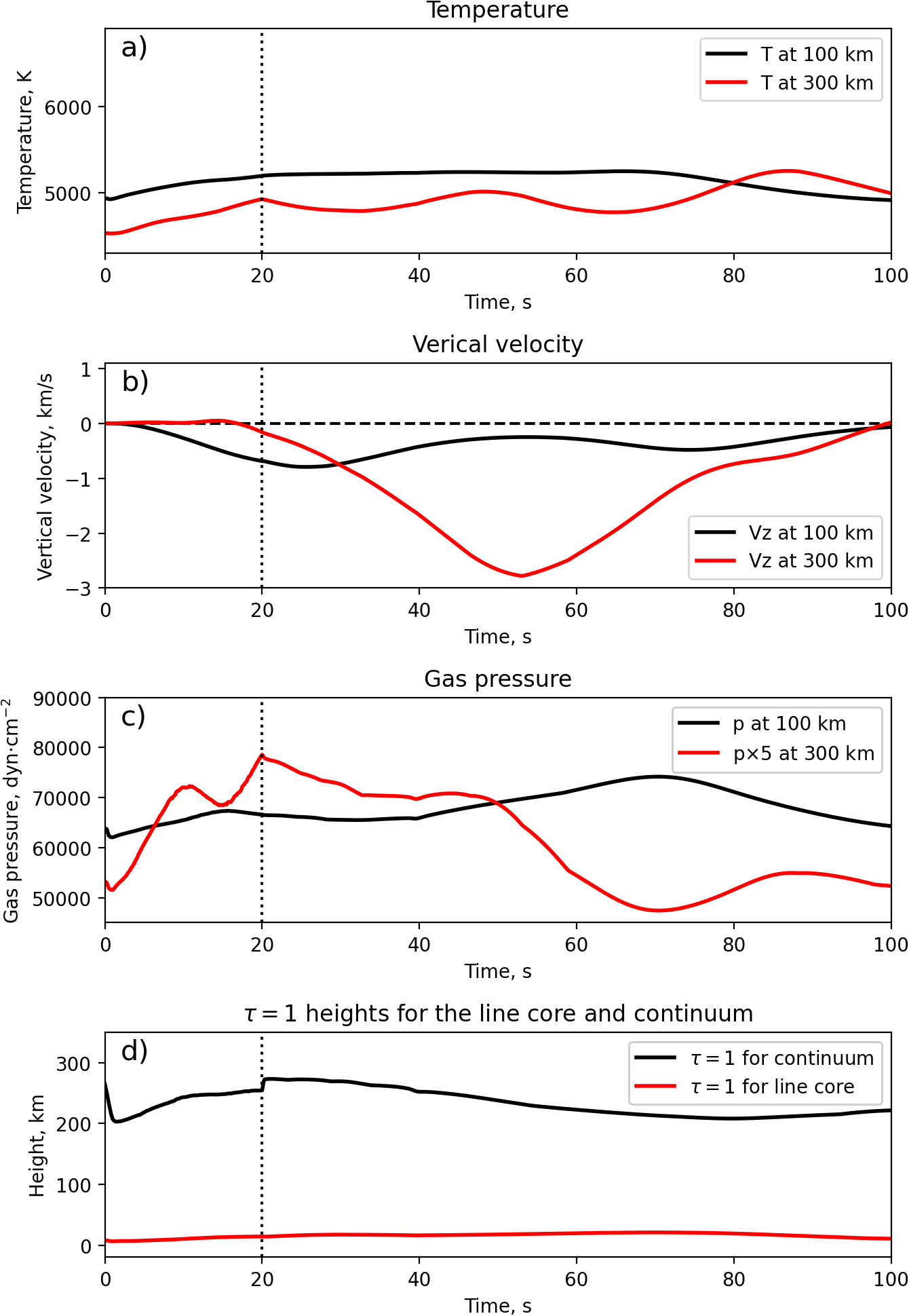}
    \includegraphics[width=0.49\linewidth]{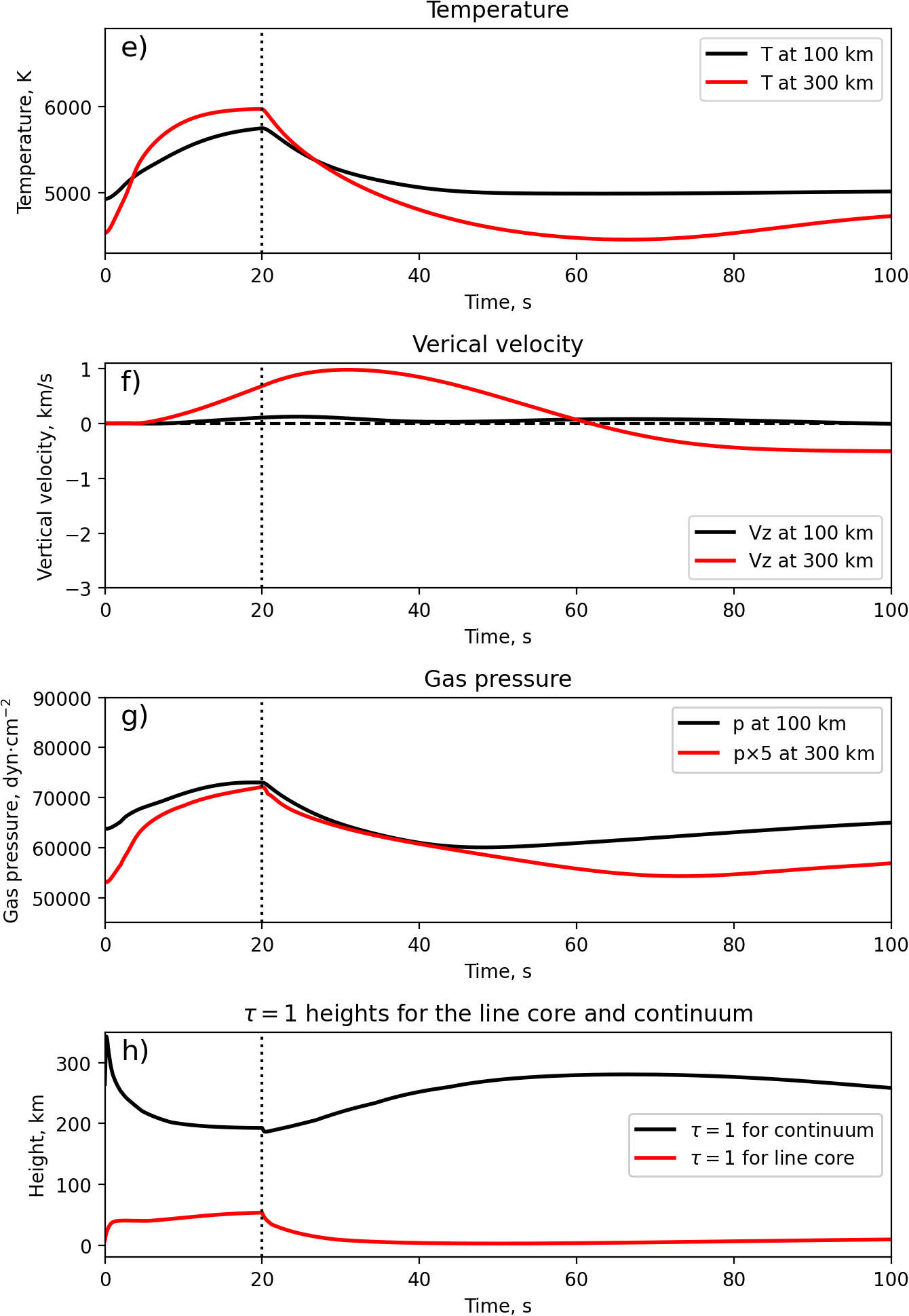}
    \caption{Illustration of the temperature (a), vertical velocity $v_{z}$ (b), and gas pressure (c) as a function of time at heights of 100~km (black solid curve) and 300~km (red solid curve) for RADYN Model 9 (E$_{c} =50$~keV, $\delta = 5$, $F_{d} = 10^{11}$ erg~cm$^{-2}$~s$^{-1}$). The $\tau = 1$ heights for the \ionwl{Fe}{1}{6173} line core and continuum are presented in panel (d). Panels (e-h) show the same parameters as in panels (a-d) but for the RADYN Model 8 ($E_c = 3000$~keV, $\delta = 3$, $F_{d} = 10^{11}$~erg~cm$^{-2}$~s$^{-1}$).}
\label{fig:atmosmodels}
\end{figure}

\begin{figure}[h]
\centering
    \includegraphics[width=1.0\linewidth]{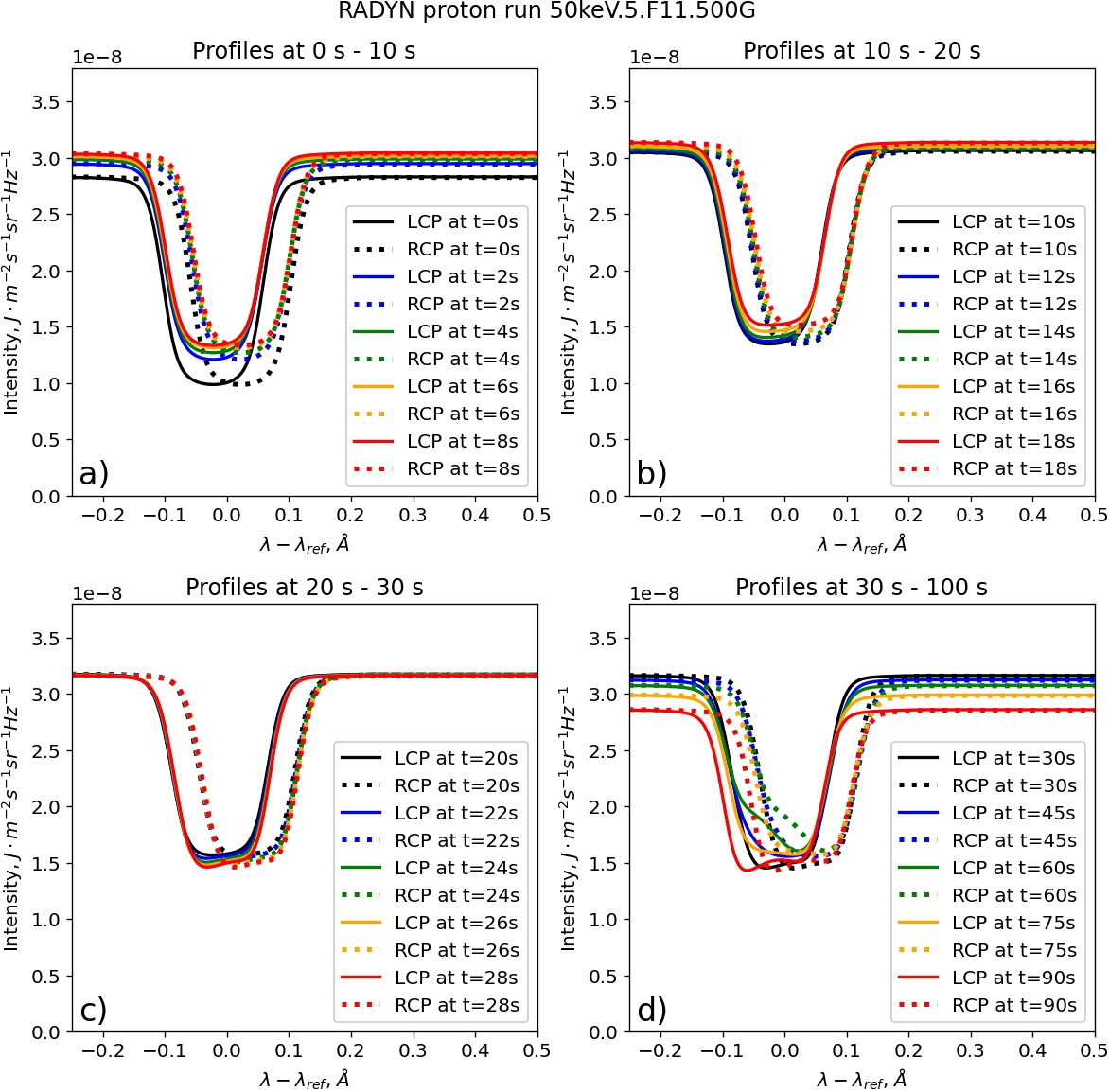}
    \caption{\ionwl{Fe}{1}{6173} LCP (solid) and RCP (dashed) line profiles for Model 9 ($E_{c} =50$~keV, $\delta =5$, $F_{d} = 10^{11}$ erg~cm$^{-2}$~s$^{-1}$) with 500~G imposed vertical magnetic field at 0--10~s (a), 10--20~s (b), 20--30~s (c), and 30--100~s (d) of the run. The times at which the profiles are sampled are coded by color, $\lambda{}_{ref}$=6173.34\,$\AA$.}
\label{fig:profiles1}
\end{figure}

\begin{figure}[h]
\centering
    \includegraphics[width=1.0\linewidth]{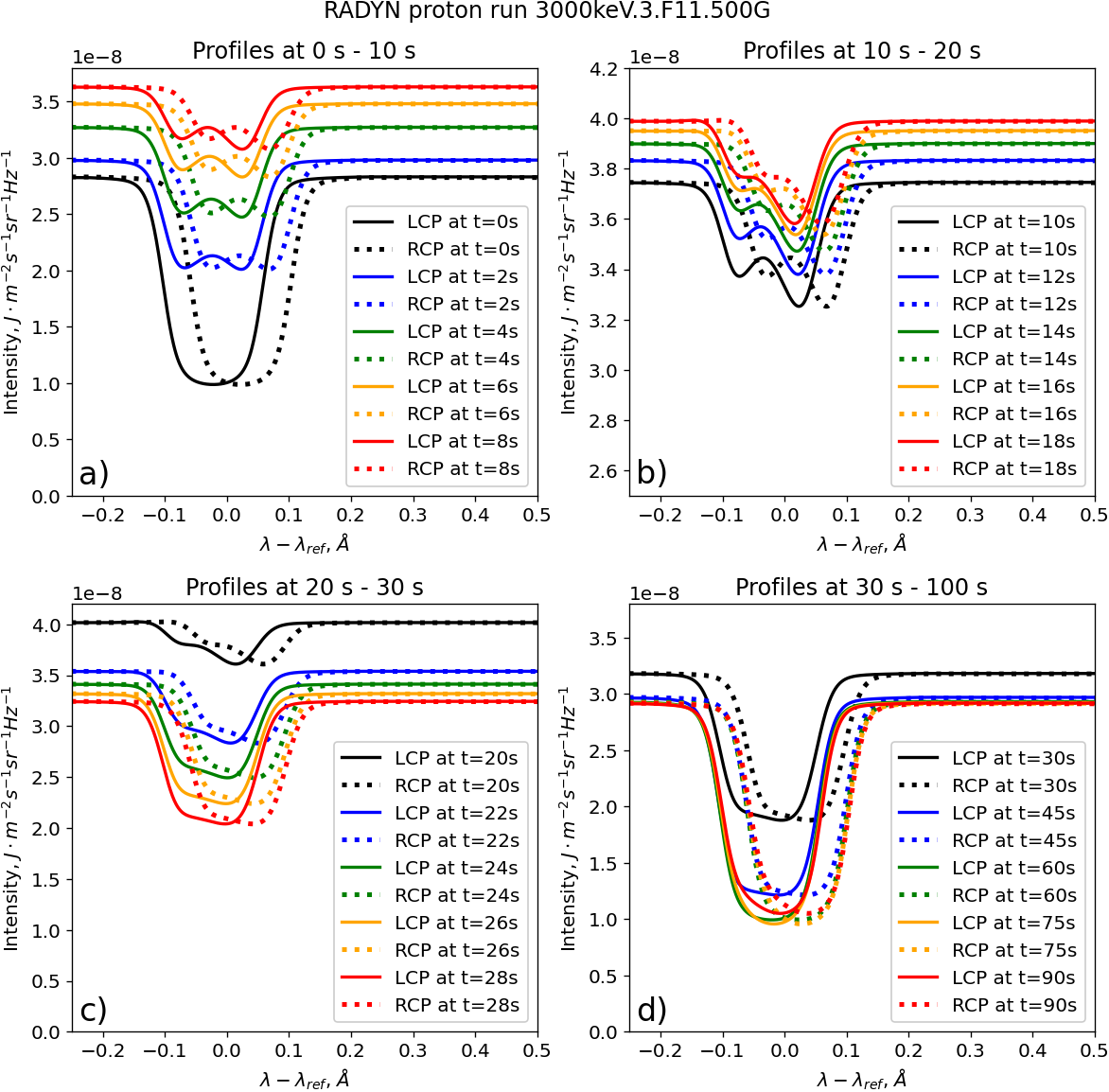}
    \caption{Same as Figure~\ref{fig:profiles1} but for Model 8 ($E_{c} = 3000$~keV, $\delta =3$, $F_{d} = 10^{11}$ erg~cm$^{-2}$~s$^{-1}$).}
    \label{fig:profiles2}
\end{figure}

\begin{figure}[h]
\centering
    \includegraphics[width=1.0\linewidth]{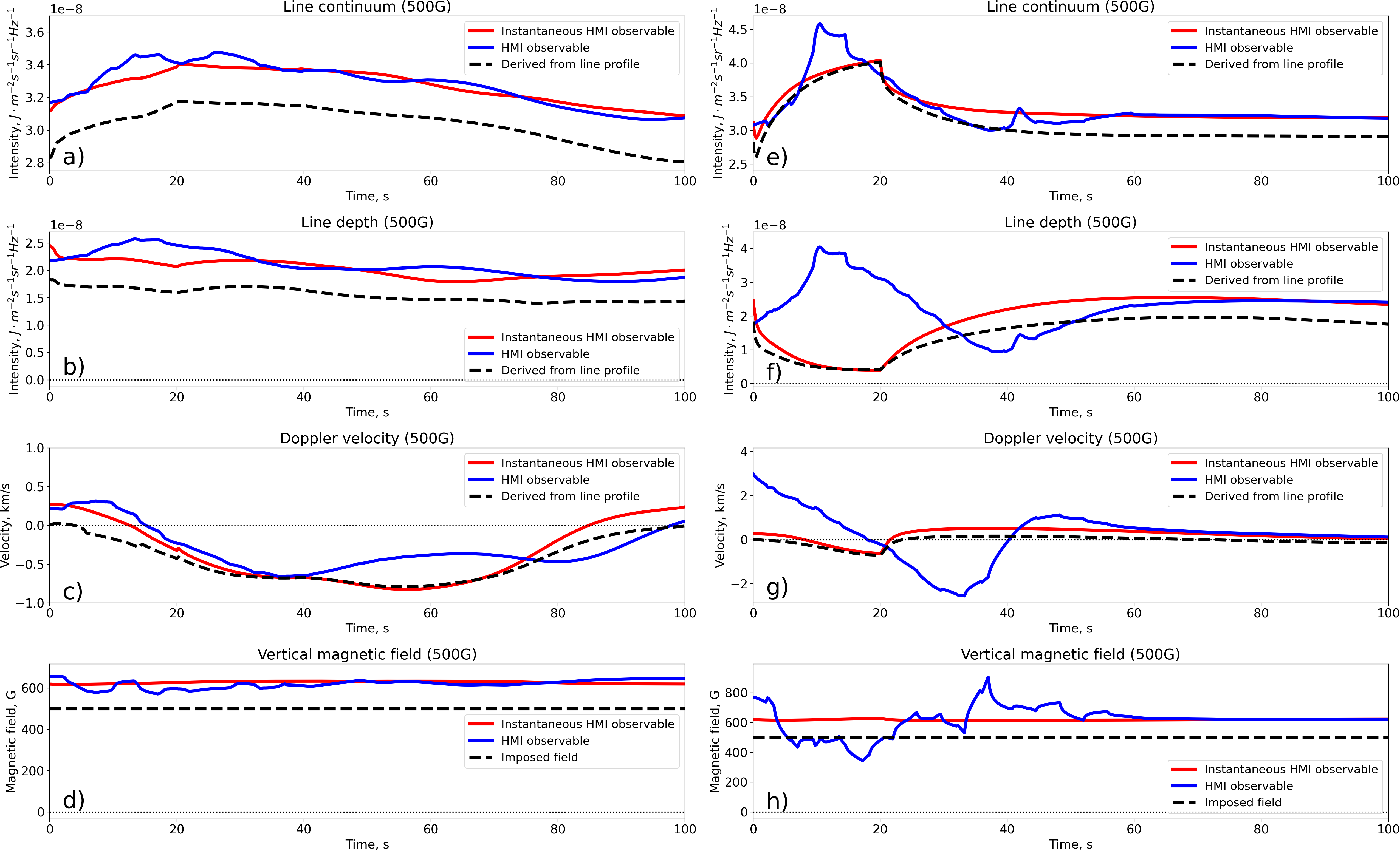}
    \caption{The continuum intensity (a), line depth (b), Doppler velocity (c), and magnetic field (d) inferred from the synthetic \ionwl{Fe}{1}{6173} line profiles from Model 9 with an imposed vertical uniform 500~G magnetic field. The black dashed curves correspond to measurements inferred from the RH-predicted line profiles (noted as $v_{D}$ in the text. The red solid curves show ``instantaneous'' observables obtained with the HMI algorithm applied to instantaneous line profiles. The blue curves show the observables obtained with the HMI algorithm applied with the actual observing sequence timing centered at the reference time. The black dot-dashed horizontal lines mark the zero level of the observables. Panels e--h illustrate the same quantities for Model 8.}
\label{fig:timedep}
\end{figure}

\begin{figure}[h]
    \centering
    \includegraphics[width=\linewidth]{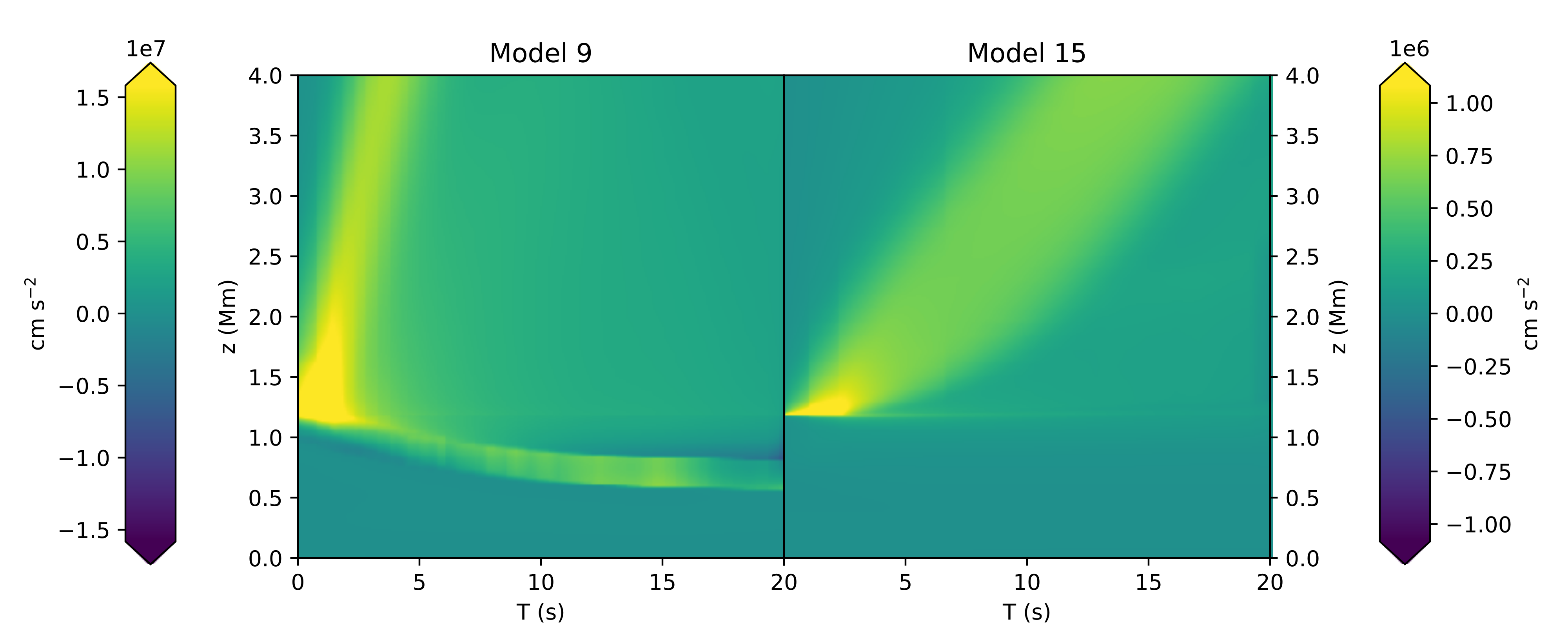}
    \caption{Time-dependent acceleration profiles derived from the RADYN proton beam simulations for Models 9 (left) and 15 (right).}
    \label{fig:acc}
\end{figure}

\begin{figure}[h]
    \centering
    \includegraphics[width=\linewidth]{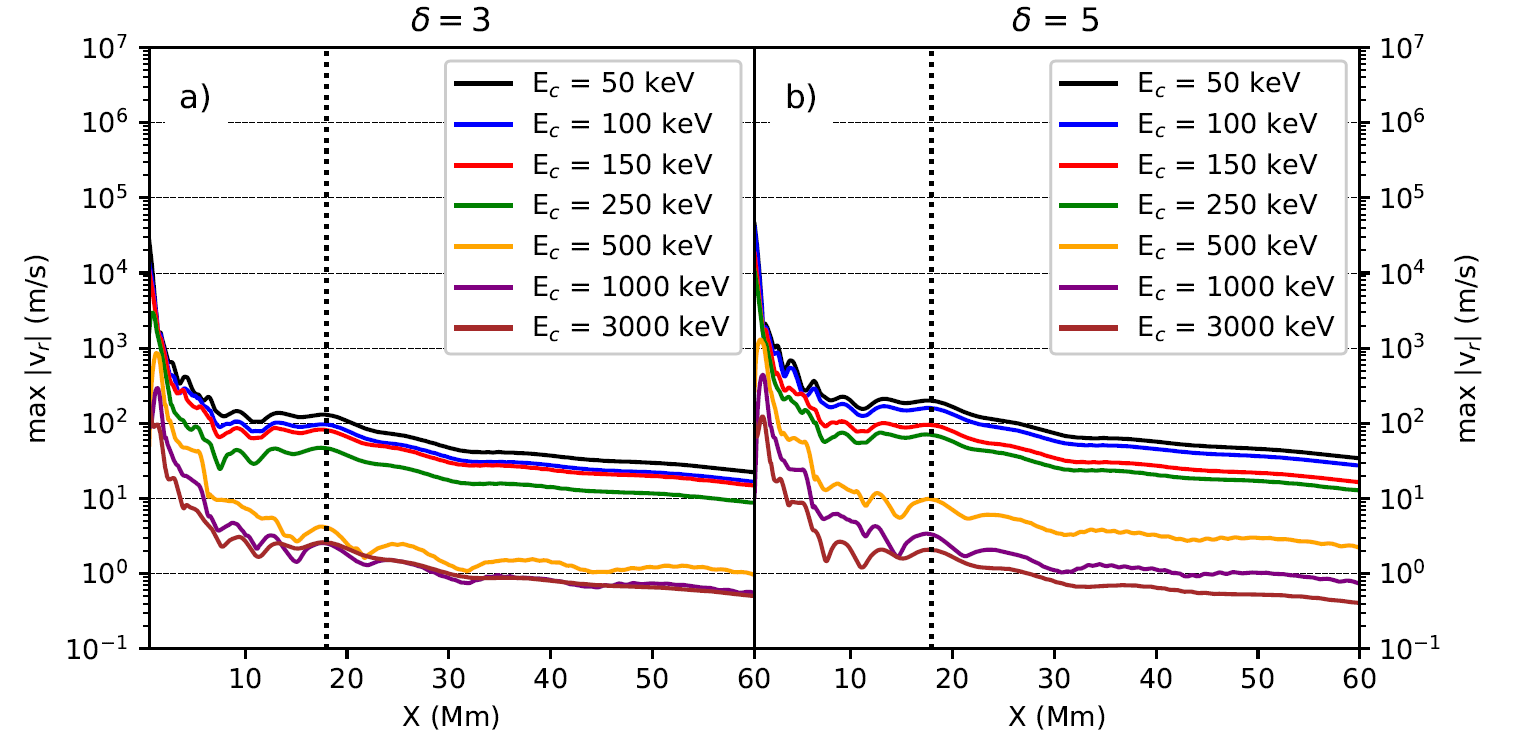}
    \caption{Log of the respective sunquake wavefront amplitudes as a function of distance from the excitation source. The typical p-mode wavefront is isolated from other waves generated in the acoustic model by considering the maximum amplitude only within five minutes of the predicted wavefront arrival time. The sample of velocities presented in Figure \ref{fig:18Mm} is marked by the vertical dashed line.}
    \label{fig:amp}
\end{figure}

\begin{figure}[h]
    \centering
    \includegraphics[width=0.5\linewidth]{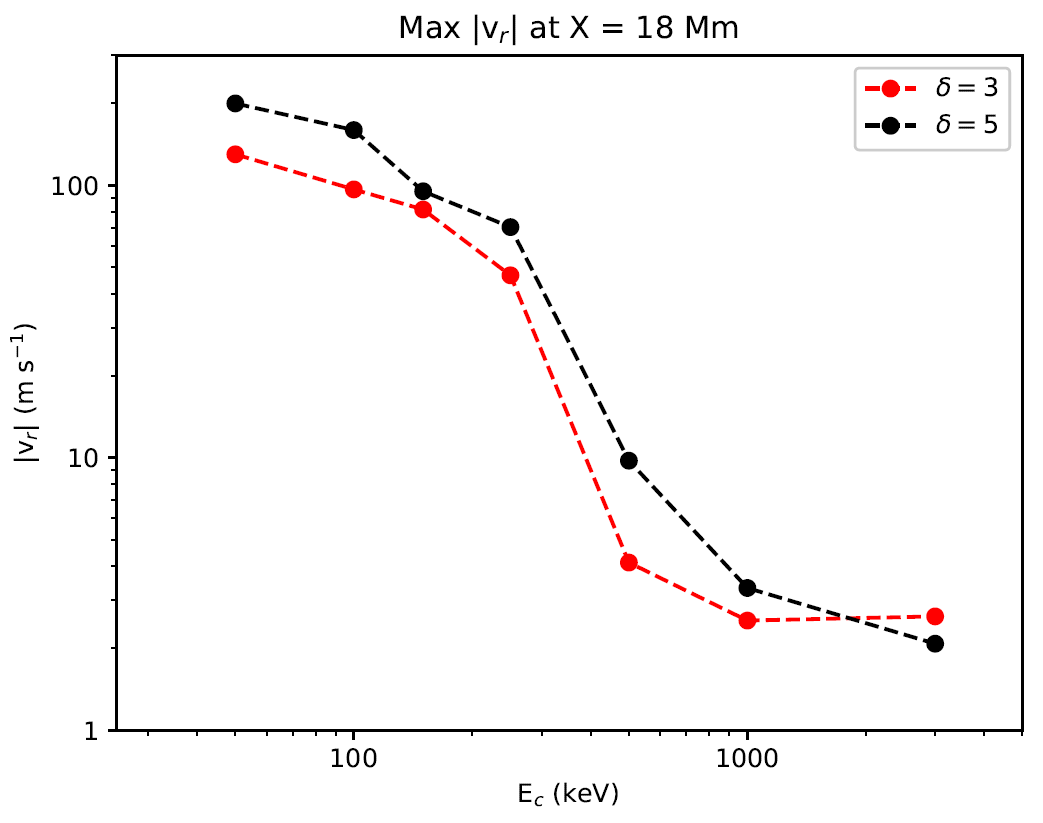}
    \caption{The sunquake wavefront amplitude at X=18 Mm for cut-off energies $E_c=50$ keV to $E_c=3000$ keV, with $\delta=3$ in red and $\delta=5$ in black.}
    \label{fig:18Mm}
\end{figure}

    To summarize the results, we claim that there are two regimes found in which the perturbations of the line profiles and the atmosphere were significant and resulted in a potential helioseismic response and/or a white-light flare (see Figure~\ref{fig:stats}). We have selected the two models, one with the strongest enhancement of the continuum near the line (Model 8, $E_{c} =3000$~keV, $\delta =3$, $F_{d} = 10^{11}$ erg~cm$^{-2}$~s$^{-1}$), and the one with the strongest Doppler shift $v_{D}$ (Model 9, $E_{c} =50$~keV, $\delta =5$, $F_{d} = 10^{11}$ erg~cm$^{-2}$~s$^{-1}$). Model 9 causes gradual but strong changes in the atmospheric parameters, resulting in $\approx$12\% increase in flare continuum during the heating phase. The Doppler velocity derived from fully-resolved polarization profiles has the strongest redshift of $v_{D}\sim{}-0.8$~km~s$^{-1}$ about 40~s after the heating. The SDO/HMI observables, in general, closely follow the properties of the full-resolution line profile. The other model, Model 8 causes a much stronger enhancement of the continuum near the \ionwl{Fe}{1}{6173} line (of the order of 42\%) with respect to Model 9 and also results in the \ionwl{Fe}{1}{6173} redshifts of $v_{D}\sim{}-0.7$~km~s$^{-1}$ at the end of the heating phase. The SDO/HMI observables in this model differ significantly from the counterparts derived from the full-resolution line profiles. Overall, both models demonstrate the impact on the deep layers of the solar atmosphere, and the hydrodynamic velocity analysis and spectral analysis of the \ionwl{Fe}{1}{6173} line do not yet allow us to claim whether the proton beams of high energy per proton or low energy per proton are the preferential candidates for causing sunquakes. To provide more insights, we utilize the responses of the atmosphere to the proton beam heating and impose this response into the 3D acoustic models (Section~\ref{sec:acousticgen}). While the Doppler velocities in the RADYN models do not differ much at either extreme of the low-energy cutoff, the physical velocities in the acoustic model are significantly stronger for Model 9 ($E_{c} =50$~keV, $\delta =5$, $F_{d} = 10^{11}$ erg~cm$^{-2}$~s$^{-1}$) than for Model 8. The continuum intensity enhancement derived from the RADYN simulation is, on the opposite, stronger for the $\delta =3$ cases, with Model 8 producing the strongest enhancement. Overall, we show not only that sunquakes can be generated from a high injected flux of protons with relatively low cutoff energy, but that these sunquakes have much higher amplitudes than those predicted from simulations with large cutoff energies. This is particularly significant as large low-energy cutoffs are more difficult to physically justify.

    The role of the proton beams in the solar flare energy deposition has been discussed for more than five decades \citep{Svestka1970,Simnett1986}, yet without a clear understanding of the energy fraction carried by these beams. Some works \citep{Emslie2012} suggest that the energy transported by the proton beams in solar flares is comparable to that of the electron beams. Proton beams are an attractive candidate as a mechanism to explain sunquake excitation as they deposit energy significantly deeper in the solar atmosphere than electron beams. To investigate the possible reasons behind the sunquake generation, we notice here that the proton beams also carry more momentum with respect to the electron beams of similar energy due to the difference in the particle masses. Given that the energy spectrum of the proton beam is determined by the power law,
    \begin{gather}
        \dfrac{dN}{dE}=
        \begin{cases}
            AE^{-\delta},    &   E\geq{}E_{c} \\
            0,  &   E<E_{c}
        \end{cases} ,
    \end{gather}
    with the constant $A$ found by normalizing to the known deposited energy rate,

    \begin{gather}
        F_{d} = \int\limits_{E_{c}}^{\infty}E\dfrac{dN}{dE}dE ~~~\Rightarrow~~~A = F_{d}\dfrac{\delta{}-2}{E_{c}^{-\delta{}+2}},
    \end{gather}
    the total momentum flux of the beam (in the non-relativistic limit) and the average energy per particle are then

    \begin{gather}
        M_{d} = \int\limits_{E_{c}}^{\infty}\sqrt{2m_{p}E}\dfrac{dN}{dE}dE = \sqrt{2m_{p}F_{d}}\dfrac{\delta{}-2}{\delta{}-\dfrac{3}{2}}\sqrt{\dfrac{F_{d}}{E_{c}}} \label{eqn:totmomentum}
        \end{gather}
        and
        \begin{gather}
        <E_{p}> = \dfrac{F_{d}}{\int\limits_{E_{c}}^{\infty}\dfrac{dN}{dE}dE} = \dfrac{\delta{}-1}{\delta{}-2}E_{c}  .\label{eqn:energyperproton}
    \end{gather}

    Although we do not analyze the momentum transport in detail in this work, it is helpful to provide some estimates. The injected momentum flux for the proton beam in Model 9 (with $F_{d} = 10^{11}$ erg~cm$^{-2}$~s$^{-1}$, $\delta =5$, and $E_{c} =50$~keV) is $M_{d} \sim 5.5 \times 10^{2}$~g cm$^{-1}$ s$^{-2}$. The total momentum deposited per unit area during 20~s therefore is $M_{d}^{tot} = M_{d} \Delta t\sim 1.1 \times 10^4$~g cm$^{-1}$ s$^{-1}$. Assuming the photospheric density $\rho\approx 10^{-7}-10^{-8}$~g cm$^{-3}$ \citep[considering the quiet Sun and sunspot model atmospheres, correspondingly;][]{Allred2015} and the characteristic scale height of $H \approx 100$~km, the bulk velocity of the plasma is $v\approx {M_{d}^{tot}}/({\rho H}) \approx 0.1-1$~km~s$^{-1}$. The latter velocity estimate is comparable to what is expected for sunquakes, and we can therefore conclude that the momentum transport by the proton beams may play an important role in sunquake initiation.

    We have explored whether proton beams of the deposited energy flux ($F_{d} = 10^{11}$ erg~cm$^{-2}$~s$^{-1}$) are capable of exciting sunquakes with amplitudes similar to observations. Depending on the frequency filtering applied to Dopplergram observations, the observed range of sunquake amplitudes measured in the line-of-sight velocity is 50--200~m s$^{-1}$ \citep{Sharykin2020,Zharkov2020}. This line-of-sight velocity is composed mainly of the radial component---reported here in Section \ref{sec:acoustic model results}---when the sunquake occurs close to the disk center. When the event occurs closer to the limb, however, the contribution from the tangential component of velocity increases; the tangential component, in general, is out of phase with the radial component and may decrease the observed sunquake amplitude at intermediate locations on the disk. For fixed cut-off energy, we find that the beam with spectral index $\delta=5$ produces a higher amplitude wavefront than the corresponding $\delta=3$ beam in nearly every case; the only outlier here is $E_c=3000$~keV, though the difference is at most only $0.5$~m s$^{-1}$. This is generally consistent with Eqn.~\ref{eqn:totmomentum} which shows that, given the same $E_{c}$ and $F_d$, larger values of $\delta$ result in larger momenta. Also, the momentum flux for fixed cut-off energy varies weakly with the spectral index, $M_{d}\propto (\delta-2)/(\delta-\frac{3}{2})$, and is only significant for smaller values of $\delta$. For example, a proton beam with $\delta = 7$ is expected to deposit only 6\% more momentum than the same beam with $\delta=5$. However, for the considered $\delta{}=3$ and $\delta{}=5$, the difference in $M_{d}$ for the beams with the same $E_{c}$ is nearly 29\%.
    
    The momentum is also inversely proportional to the $\sqrt{E_{c}}$ which explains the increasing trend of the Doppler velocities in Fig.~\ref{fig:stats}b. Decreasing the cut-off energy has a similar effect on the energy spectrum of the proton beam as increasing the spectral index, though the momentum flux depends more strongly on the cut-off energy, $M\propto1/\sqrt{E_{c}}$. This dependence is only noticeable for $E_c \ge 250$~keV; for cut-off energies less than this, the decrease in sunquake amplitude with increasing cut-off energy is much milder. This weak dependence on cut-off energy in the lower end of the parameter space is measurable in principle by HMI with its precision of 13~m s$^{-1}$ at disk center \citep{schou2012HMImeasurements}, though in practice the addition of the background convective noise would make this difference very difficult to measure. Furthermore, different input beam energy fluxes in observed sunquakes may further convolute the amplitude discrepancy.

\begin{sidewaysfigure}[h]
\centering
\includegraphics[width=1.0\linewidth]{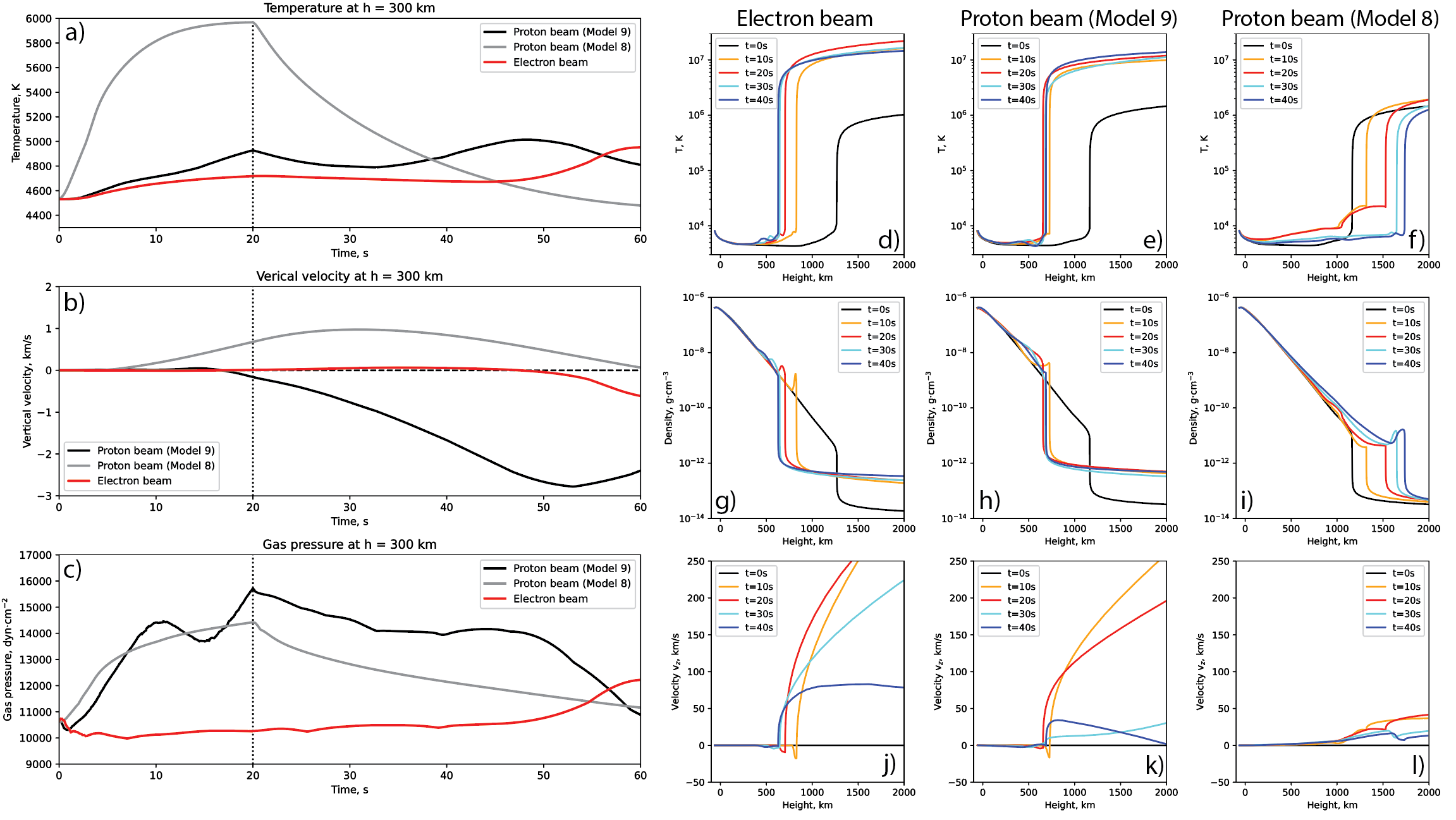}
\caption{Illustration of the temperature (a), vertical velocity $v_{z}$ (b), and gas pressure (c) as a function of time at the height of 300~km for Model 9 (black line), Model 8 (gray line), and for a RADYN simulation of non-thermal electron beam heating characterized by a power-law with $E_{c} =15$~keV, $\delta =5$, $F_{d} = 10^{11}$ erg~cm$^{-2}$~s$^{-1}$ \citep[considered in][red line]{Graham2020}. Panels (d-f) highlight the evolution of the temperature profiles for these models, panels (g-i)~--- densities, and panels (j-l)~--- vertical velocities.}
\label{fig:protons_electrons}
\end{sidewaysfigure}

    The F-CHROMA models considered in~\citet{sadykov2020FCHROMAruns} have a peak energy deposition rate of $F_{d} = 10^{11}$ erg~cm$^{-2}$~s$^{-1}$. However, the average energy deposition rate was twice lower than that value, which may impact the close comparison of F-CHROMA models and the RADYN proton beam heating models considered in this work. Figure~\ref{fig:protons_electrons} illustrates a comparison of the response of the atmospheric parameters at $h = 300$~km for the proton beam model $F_{d} = 10^{11}$ erg~cm$^{-2}$~s$^{-1}$, $E_c=50$~keV, $\delta=5$ (Model 9) and the electron beam heating model with $F_{d} = 10^{11}$ erg~cm$^{-2}$~s$^{-1}$, $E_c=15$~keV, $\delta=5$ from \citet{Graham2020}. The RADYN electron beam heating model had a duration of heating of 20~s similar to the one in proton models, and a total duration of the run of 60~s. One can see in Figure~\ref{fig:protons_electrons}a-c that while experiencing weaker but still comparable temperature enhancement, the electron beam heating model does not result in strong downward velocities and pressure enhancements at $h = 300$~km. The dynamics of the atmospheres for the considered models are presented in more detail in Figure~\ref{fig:protons_electrons}d-l. While we do not perform a detailed study of the atmospheric dynamics in this work, it is worth noting that the behavior of the proton beam-header atmosphere for the beam with a large $E_{c}=3000\,keV$ proton beam model (Model 8) differs dramatically from the behavior of the small $E_{c}=50\,keV$ proton beam model (Model 9), as well as from the considered electron beam heating model.

    The correlation of the white light emission (including SDO/HMI observable continuum) with the hard X-ray sources in solar flares  \citep{Watanabe2020,Battaglia2012} suggests that the electron beam heating and radiative back-warming can enhance the white light emission. Fig.~\ref{fig:stats}a demonstrates that the proton beams can contribute to the enhancements of the continuum near the \ionwl{Fe}{1}{6173} line as well. The Figure also shows a tendency of the white-light enhancements to increase with the $E_{c}$ increase and preferring lower $\delta$ values. This is in qualitative agreement with Eqn.~\ref{eqn:energyperproton} which demonstrates that (1) for the same $E_{c}$ the $\delta{}=3$ beam particles have, on average, 50\% more energy per particle, and (2) the $<E_{p}>$ is directly proportional to $E_{c}$. The higher values of $<E_{p}>$ would indicate that the proton beam particles, on average, have to encounter thicker media to lose their energy through thermalization and, therefore, have to penetrate deeper into the solar atmosphere. All of the considered proton beam models exceeded this enhancement: 3.9\% was the weakest enhancement value observed for Models 11 and 12, Models 6-9 and 15 had an enhancement of more than 10\%, and the remaining Model enhancements were concentrated within the 6\%-10\% range. It is also important to notice that the enhancements reported in this work are derived directly from the computed continua near the \ionwl{Fe}{1}{6173} line and not after applying the SDO/HMI observable algorithm, which can potentially generate artificial enhancements of the continua \citep{Mravcova2017,Svanda2018}.

    For the models considered in this work, we have found two separate regimes in the low-energy cutoff spectrum for the proton beams: small $E_{c}$ beams which produce a strong seismic signal, and large $E_{c}$ beams which produce comparatively more intense continuum emission. Observational evidence for such a duality can be found in, for example, the analysis of \citet{Pedram} where white light enhancement for a sample of nine flares was found to be noticeably weaker in sunquake-generating events. However, first, both the RADYN model and acoustic model show an increase in the Doppler velocity and physical velocity, respectively, in the $E_c=3000$~keV beam when the spectral index steepens from $\delta = 5$ to $\delta = 3$. It is possible that such large low-energy cutoffs can explain sunquake-generating events with significant white-light enhancement, though an extension of this analysis to higher low-energy cutoffs and lower spectral indices is necessary to verify the trend. Second, \ionwl{Fe}{1}{6173} line profiles for Model 9 that resulted in a strong helioseismic response in 3D acoustic simulations (see Figure~\ref{fig:profiles1}) disagree with the observations by SDO/HMI of at least some sunquake events. Specifically, the filtergrams presented by \citet[][see Figure3b]{Sharykin2020} and, more recently, by \citet[][see Figure 10]{Kosovichev2023} demonstrate that the absorption feature of the \ionwl{Fe}{1}{6173} line almost disappears at the location of the sunquake photospheric source. Such behavior qualitatively agrees with the profiles synthesized for Model 8 (see Figure~\ref{fig:profiles2}) but not for Model 9. Given that weaker FCHROMA electron beams are found to suppress the \ionwl{Fe}{1}{6173} line absorption feature by $\sim$30\% \citep[see Figure 2 in][]{sadykov2020FCHROMAruns}, the consideration of the impact by both the proton and electron beams together on the atmosphere may be necessary to explain the line profile dynamics.

    A population of the very high-energy ($>$30~MeV) protons can be diagnosed by producing the 2.223~MeV neutron-capture $\gamma$-ray line \citep{Shih2009}. However, the lower-energy proton distribution is currently almost `invisible' to the observer. Recently, \citet{Kerr2023} studied the Orrall-Zirker effect~\citep{Orrall1976} by performing proton beam driven RADYN simulations, with $F_{d}$=10$^{9}$-10$^{11}$~erg~cm$^{-2}$~s$^{-1}$, $E_c=150$~keV, $\delta=5$. Though their models predicted a much weaker, and very transient, signal than earlier experiments suggested, \cite{Kerr2023} did find a detectable non-thermal enhancement of Lyman lines produced via a charge exchange between the protons in the beam and the ambient plasma and indicated the potential possibility to diagnose the injected beam via $Ly\beta$ observations from the Spectral Imaging of the Coronal Environment (SPICE) Instrument onboard the Solar Orbiter. Figure~4 of \citep{Orrall1976} also indicates that the enhancements of the $Ly\alpha$ wings are most notable for lower $\delta$ values and for the $\approx$30~keV protons \citep[also noticed in][]{Simnett1995}. This provides an opportunity to observationally test the modeling-based selection rule that lower-energy proton beams are responsible for sunquakes, especially given that many strong flares have helioseismic counterparts \citep{Sharykin2020}.

\section*{\textsc{Acknowledgments}} \noindent \centering \normalsize This research was supported by NSF grants 1916509 and 1835958 and NASA grants NNX14AB68G and NNX16AP05H. VMS acknowledges the NSF FDSS grant 1936361. GSK acknowledges the financial support from a NASA Early Career Investigator Program award (Grant\# 80NSSC21K0460). JCA acknowledges funding from the Heliophysics Innovation Fund of the ISFM program and from the Heliophysics Supporting Research Program. AFK acknowledges the NSF grant 1916511.

\newpage
\begin{longrotatetable}
\begin{deluxetable}{clcccccccccc}
\tablecaption{Properties of the atmosphere and the \ionwl{Fe}{1}{6173} line formation for RADYN proton beam simulations with $\delta =3$, $F_{d} = 10^{11}$ erg~cm$^{-2}$~s$^{-1}$. The subscript '$0$' denotes the parameter value at t=0~s. The '$min$' and '$max$' operators return the minimum and the maximum values of the parameters during the run. $I_c$ denotes the continuum near the line, $I_d$~--- line depth (measured as a difference between the line continuum and the smallest intensity across the line), $\tau{}_{I_c}$~--- the optical depth for the continuum near the \ionwl{Fe}{1}{6173} line, and $\tau{}_{I_d}$~--- the optical depth for the line core.}
\tablewidth{0pt}
\tablehead{\colhead{Model} & \colhead{Model properties and height}  &   \colhead{$\dfrac{max(T)}{T_0}$}   &   \colhead{$\dfrac{max(p)}{p_0}$}   &   \colhead{max(v$_z$)}    &   \colhead{min(v$_z$)}    &   
\colhead{$\dfrac{max(I_c)}{I_{c0}}$} &   \colhead{$\dfrac{max(I_d)}{I_{c0}}$}   &   \colhead{max(v$_{D}$)}  &   \colhead{min(v$_{D}$)}  &   \colhead{max($\tau{}_{I_c}$=1)}  &    \colhead{max($\tau{}_{I_d}$=1)} \\
\colhead{}    &   \colhead{}    &    \colhead{}  &    \colhead{}  &   \colhead{(km~s$^{-1}$)}  &   \colhead{(km~s$^{-1}$)}  &  \colhead{} & \colhead{}  &  \colhead{(km~s$^{-1}$)}  &  \colhead{(km~s$^{-1}$)}  &  \colhead{(km)}   &   \colhead{(km)}
}
\startdata
1 & $E_{c} = 25$~keV, $\delta =3$, $h = 100$~km & 1.043 & 1.089 & 0.000 & -0.517 & 1.077 & 0.497 & 0.022 & -0.307 & 15 & 265 \\
& ~---, $h = 300$~km & 1.125 & 1.267 & 0.338 & -1.725 &  ~--- & ~--- & ~--- & ~--- &  ~--- & ~--- \\
\hline
2 & $E_{c} =50$~keV, $\delta =3$, $h = 100$~km & 1.041 & 1.068 & 0.025 & -0.368 & 1.073 & 0.513 & 0.034 & -0.243 & 15 & 265 \\
& ~---, $h = 300$~km & 1.109 & 1.244 & 0.523 & -1.349 &  ~--- & ~--- & ~--- & ~--- &  ~--- & ~--- \\
\hline
3 & $E_{c} = 100$~keV, $\delta =3$, $h = 100$~km & 1.041 & 1.041 & 0.011 & -0.162 & 1.070 & 0.517 & 0.168 & -0.132 & 15 & 265 \\
& ~---, $h = 300$~km & 1.068 & 1.163 & 0.271 & -0.717 &  ~--- & ~--- & ~--- & ~--- &  ~--- & ~--- \\
\hline
4 & $E_{c} = 150$~keV, $\delta =3$,$h = 100$~km & 1.052 & 1.037 & 0.028 & -0.099 & 1.082 & 0.488 & 0.150 & -0.081 & 16 & 265 \\
& ~---, $h = 300$~km & 1.072 & 1.109 & 0.225 & -0.453 &  ~--- & ~--- & ~--- & ~--- &  ~--- & ~--- \\
\hline
5 & $E_{c} = 250$~keV, $\delta =3$, $h = 100$~km & 1.057 & 1.038 & 0.033 & -0.021 & 1.090 & 0.456 & 0.117 & -0.032 & 17 & 265 \\
& ~---, $h = 300$~km & 1.086 & 1.109 & 0.244 & -0.289 &  ~--- & ~--- & ~--- & ~--- &  ~--- & ~--- \\
\hline
6 & $E_{c} =500$~keV, $\delta =3$, $h = 100$~km & 1.077 & 1.063 & 0.026 & -0.271 & 1.155 & 0.376 & 0.025 & -0.129 & 22 & 271 \\
& ~---, $h = 300$~km & 1.156 & 1.328 & 0.316 & -0.363 &  ~--- & ~--- & ~--- & ~--- &  ~--- & ~--- \\
\hline
7 & $E_{c} = 1000$~keV, $\delta =3$, $h = 100$~km & 1.087 & 1.067 & 0.045 & 0.000 & 1.174 & 0.303 & 0.082 & -0.091 & 26 & 271 \\
& ~---, $h = 300$~km & 1.195 & 1.211 & 0.541 & -0.317 &  ~--- & ~--- & ~--- & ~--- &  ~--- & ~--- \\
\hline
8 & $E_{c} = 3000$~keV, $\delta =3$, $h = 100$~km & 1.166 & 1.143 & 0.121 & -0.011 & 1.421 & 0.141 & 0.167 & -0.686 & 53 & 343 \\
& ~---, $h = 300$~km & 1.316 & 1.354 & 0.975 & -0.509 &  ~--- & ~--- & ~--- & ~--- &  ~--- & ~--- \\
\hline
\enddata
\label{tab:stats_f11_d3}
\end{deluxetable}
\end{longrotatetable}

\newpage
\begin{longrotatetable}
\begin{deluxetable}{clcccccccccc}
\tablecaption{Same as Table~\ref{tab:stats_f11_d3} but for the models with $\delta =5$, $F_{d} = 10^{11}$ erg~cm$^{-2}$~s$^{-1}$.}
\tablewidth{0pt}
\tablehead{\colhead{Model}  & \colhead{Model properties and height}  &   \colhead{$\dfrac{max(T)}{T_0}$}   &   \colhead{$\dfrac{max(p)}{p_0}$}   &   \colhead{max(v$_z$)}    &   \colhead{min(v$_z$)}    &   
\colhead{$\dfrac{max(I_c)}{I_{c0}}$} &   \colhead{$\dfrac{max(I_d)}{I_{c0}}$}   &   \colhead{max(v$_{D}$)}  &   \colhead{min(v$_{D}$)}  &   \colhead{max($\tau{}_{I_c}$=1)}  &    \colhead{max($\tau{}_{I_d}$=1)} \\
\colhead{}    &   \colhead{}    &   \colhead{} &    \colhead{}   &   \colhead{(km~s$^{-1}$)}  &   \colhead{(km~s$^{-1}$)}  &  \colhead{} & \colhead{}  &  \colhead{(km~s$^{-1}$)}  &  \colhead{(km~s$^{-1}$)}  &  \colhead{(km)}   &   \colhead{(km)}
}
\startdata
9  & $E_{c} =50$~keV, $\delta =5$, $h = 100$~km & 1.065 & 1.161 & 0.000 & -0.794 & 1.122 & 0.493 & 0.026 & -0.796 & 21 & 273 \\
& ~---, $h = 300$~km & 1.159 & 1.476 & 0.048 & -2.779 &  ~--- & ~--- & ~--- & ~--- &  ~--- & ~--- \\
\hline
10 & $E_{c} = 100$~keV, $\delta =5$, $h = 100$~km & 1.055 & 1.125 & 0.000 & -0.650 & 1.087 & 0.509 & 0.023 & -0.492 & 17 & 265 \\
& ~---, $h = 300$~km & 1.129 & 1.302 & 0.256 & -2.121 &  ~--- & ~--- & ~--- & ~--- &  ~--- & ~--- \\
\hline
11 & $E_{c} = 150$~keV, $\delta =5$, $h = 100$~km & 1.023 & 1.036 & 0.005 & -0.191 & 1.039 & 0.540 & 0.127 & -0.146 & 10 & 265 \\
& ~---, $h = 300$~km & 1.089 & 1.188 & 0.191 & -0.857 &  ~--- & ~--- & ~--- & ~--- &  ~--- & ~--- \\
\hline
12 & $E_{c} = 250$~keV, $\delta =5$, $h = 100$~km & 1.019 & 1.014 & 0.008 & -0.109 & 1.039 & 0.581 & 0.101 & -0.068 & 9 & 265 \\
& ~---, $h = 300$~km & 1.052 & 1.102 & 0.082 & -0.491 &  ~--- & ~--- & ~--- & ~--- &  ~--- & ~--- \\
\hline
13 & $E_{c} =500$~keV, $\delta =5$, $h = 100$~km & 1.037 & 1.026 & 0.002 & -0.172 & 1.069 & 0.584 & 0.025 & -0.125 & 11 & 265 \\
& ~---, $h = 300$~km & 1.038 & 1.110 & 0.069 & -0.366 &  ~--- & ~--- & ~--- & ~--- &  ~--- & ~--- \\
\hline
14 & $E_{c} = 1000$~keV, $\delta =5$, $h = 100$~km & 1.039 & 1.022 & 0.021 & -0.012 & 1.065 & 0.549 & 0.065 & -0.005 & 11 & 265 \\
& ~---, $h = 300$~km & 1.050 & 1.054 & 0.177 & -0.109 &  ~--- & ~--- & ~--- & ~--- &  ~--- & ~--- \\
\hline
15 & $E_{c} = 3000$~keV, $\delta =5$, $h = 100$~km & 1.130 & 1.105 & 0.121 & -0.008 & 1.346 & 0.260 & 0.126 & -0.094 & 30 & 276 \\
& ~---, $h = 300$~km & 1.231 & 1.292 & 0.783 & -0.379 &  ~--- & ~--- & ~--- & ~--- &  ~--- & ~--- \\
\hline
\enddata
\label{tab:stats_f11_d5}
\end{deluxetable}
\end{longrotatetable}

\newpage

\newpage

\newpage

\newpage

\newpage

\newpage

\newpage

\newpage

\newpage

\newpage

\newpage

\bibliography{radynprotons_sunquakes}{}
\bibliographystyle{aasjournal}

\end{document}